\documentclass[a4paper,fleqn,usenatbib]{mnras}

\usepackage{mathptmx}
\usepackage[T1]{fontenc}
\usepackage{ae,aecompl}

\usepackage{graphicx}	
\usepackage{amsmath}	
\usepackage{amssymb}	
\usepackage{pdflscape}	

\title[The 3XMM/SDSS Stripe 82 Galaxy Cluster Survey. II]{The 3XMM/SDSS Stripe 82 Galaxy Cluster Survey\\ II. X-ray and optical properties of the cluster sample}

\author[Takey et al. ]{
        Ali Takey,$^{1}$\thanks{E-mail: ali.takey@nriag.sci.eg} 
        Florence Durret,$^{2}$
        Isabel M\'arquez,$^{3}$
        Amael Ellien,$^{2}$
        Mona Molham,$^{1}$
\newauthor and Ad\`ele Plat$^{2}$
        \\
        $^{1}$National Research Institute of Astronomy and Geophysics (NRIAG), 
        11421 Helwan, Cairo, Egypt\\
        $^{2}$Sorbonne Universit\'es, UPMC Univ. Paris 6 et CNRS, UMR~7095, 
              Institut d'Astrophysique de Paris (IAP), \\
              98bis Bd Arago, 75014 Paris, France\\             
        $^{3}$Instituto de Astrof\'isica de Andaluc\'ia (IAA-CSIC), E-18008 Granada, Spain
}

\date{Accepted XXX. Received YYY; in original form ZZZ}

\pubyear{2018}

\begin{document}
\label{firstpage}
\pagerange{\pageref{firstpage}--\pageref{lastpage}}
\maketitle


\begin{abstract}
  We present X-ray and optical properties of the optically confirmed
  galaxy cluster sample from the 3XMM/SDSS Stripe 82 cluster
  survey. The sample includes 54 galaxy clusters in the redshift range
  of 0.05-1.2, with a median redshift of 0.36. We first present the
  X-ray temperature and luminosity measurements that are used to
  investigate the X-ray luminosity-temperature relation. The slope and
  intercept of the relation are consistent with those published in the
  literature. Then, we investigate the optical properties of the
  cluster galaxies including their morphological analysis and the
  galaxy luminosity functions. The morphological content of cluster
  galaxies is investigated as a function of cluster mass and distance
  from the cluster center. No strong variation of the fraction of
  early and late type galaxies with cluster mass is observed. The
  fraction of early type galaxies as a function of cluster radius
  varies as expected. The individual galaxy luminosity
  functions (GLFs) of red sequence galaxies were studied in the five
  ugriz bands for 48 clusters. The GLFs were then stacked in three mass 
  bins and two redshift bins. Twenty clusters of the present sample
  are studied for the first time in X-rays, and all are studied for
  the first time in the optical range. Altogether, our sample appears
  to have X-ray and optical properties typical of ``average'' cluster
  properties.

\end{abstract}





\section{Introduction}

Galaxy clusters are the largest massive structures in the universe which contain hundreds to thousands of galaxies within spatial regions of a few Mpc. They also contain gas in their intracluster medium (ICM) that is smoothly distributed and filling the intergalactic space. The hot ICM is a key feature in studying galaxy clusters since it is a strong X-ray emitter, which allows the identification process up to high redshifts and reveals that clusters are well defined and connected structural entities. The study of galaxy clusters gives the opportunity to investigate the physical processes behind the formation and evolution of their baryonic components (galaxies and gas) and to probe the distribution of matter in the universe \citep[e.g. ][]{Boehringer08, Allen11}. 

The X-ray selection of galaxy clusters has several advantages for cosmological studies: the observable X-ray temperature and luminosity of a cluster is tightly correlated with the cluster total mass, which is the most fundamental parameter for clusters \citep{Reiprich02}. Also, the cluster X-ray luminosity correlates well with its temperature (${\rm L_X-T_X}$), following the relation predicted by cluster formation models. For example, the self-similar model \citep{Kaiser86} simply predicts that clusters formed by gravitational collapse in the universe and that massive galaxy clusters are a scaled version of small clusters. Hence, cluster masses can be inferred from scaling relations found between cluster observable properties. 
Many studies stated that the slope of the ${\rm L_X-T_X}$ relation is steeper than that expected from a self-similar model (${\rm L_X \propto T_X^2}$) for various samples of galaxy groups and clusters \citep[e.g.][]{Pratt09,Hilton12,Takey13,Giles16}. It is also important to track the  ${\rm L_X-T_X}$ relation  with redshift since different heating mechanisms can be involved. 

Galaxy clusters are also considered as the largest astrophysical laboratories that are suitable to investigate the galaxy formation, evolution, and morphological properties within a well defined dense environment. 
This environment is known to influence galaxy properties. 
Morphological segregation of galaxies in clusters was indeed found to
be strong since the seminal paper of \cite{Dressler80}, who showed for a sample of 55 nearby clusters ($z<0.07$) that early-type galaxies were dominant in the central regions of clusters while late-type galaxies were more abundant in the outskirts. It was later shown to be also the case in more distant clusters and explained by the fact that numerous galaxy mergers take place in cluster centers, thus creating a large population of early-type galaxies, while late-type galaxies are continuously accreted from the field onto clusters along the cosmic filaments at the intersection of which clusters are believed to be located \citep[e.g. ][]{Adami09}.
Galaxy luminosity functions (GLFs) have also been found to depend on the
environment, with a difference between cluster and field
galaxies, and a flattening of the GLF as the environment becomes less
dense (as described in detail in \cite{Martinet15} and references
therein).  

In a previous paper on galaxy clusters in the SDSS Stripe 82, \citet{Durret15} investigated the fraction of late-type to early-type galaxies with cluster redshift. They also investigated the evolution of the galaxy luminosity function with redshift. This study was based on cluster candidates with only photometric redshifts extracted from the SDSS Stripe 82 (S82, hereafter) data. 

In the present paper, we investigate the above mentioned studies (${\rm L_X-T_X}$ relation, morphological analysis, and galaxy luminosity function) for the galaxy cluster sample conducted in the cluster survey published by \citet{Takey16}. The cluster sample includes X-ray selected and optically confirmed clusters from XMM-Newton and S82 data, respectively. We will first investigate the relation between the X-ray luminosity and temperature of the cluster sample, which spans a wide redshift range. We will then study the morphology and luminosity function of cluster galaxies in our sample as a function of cluster properties (cluster mass and redshift).     

The paper structure is as follows. We first present in Sect. 2 the cluster sample used in our analysis. X-ray data reduction and analysis as well as the ${\rm L_X-T_X}$ relation are presented in Sect. 3. The morphological properties and galaxy luminosity functions of cluster galaxies are presented in Sects. 4 and 5, respectively. We finally summarise our work and conclude in Sect. 6. 
We use the cosmological constants 
$\Omega_{\rm M}=0.3$, $\Omega_{\Lambda}=0.7$ and $H_0=70$\ km\ s$^{-1}$\ Mpc$^{-1}$ throughout the paper.



\section{The galaxy cluster sample}

We have published a galaxy group/cluster catalogue in the framework of the 3XMM/SDSS Stripe 82 galaxy cluster survey \citep{Takey16}. The survey was based on X-ray extended sources from the third XMM-Newton serendipitous source catalogue \citep[3XMM-DR5,][]{Rosen15} that are located on the sky coverage of the SDSS S82. The survey area is 11.25 deg$^2$ due to the relatively small number of XMM-Newton observations (74 pointings) targeting celestial objects and/or positions in the S82 footprint. We limited the cluster search to sources located in the S82 region, where the optical data are deeper than in the normal SDSS survey.  These 74 observations span a wide range of exposure times (good time intervals) from 2~ks to 65~ks. Also, these observations are clean ones that have only a masked area $\leq 1 \%$. The masked areas are not suitable for source detections. We also required that at least one of the EPIC cameras is used in full frame mode, so that the full field of view is exposed.

We then selected all the X-ray extended sources from the 3XMM-DR5 catalogue that are detected in the EPIC images of the 74 observations considered in our cluster survey. This list includes 120 detections that contain multiple and spurious detections. By avoiding the multiplicity and removing possible spurious detections through visual inspection of their X-ray and optical images, the X-ray galaxy cluster candidate list comprises 94 extended sources. By cross-matching this list with six X-ray and optically selected cluster catalogues and by searching the NASA/IPAC Extragalactic Database (NED), we constructed a cluster catalogue comprising 54 galaxy clusters that are known in the literature with measured redshifts. The remaining candidates (40 sources) have no redshifts in the literature and are expected to be distant clusters. The list of the galaxy cluster catalogue (54 clusters) and the 40 X-ray cluster candidates are published in our first paper by \citet{Takey16}.     

\begin{figure}
\includegraphics[width=\columnwidth]{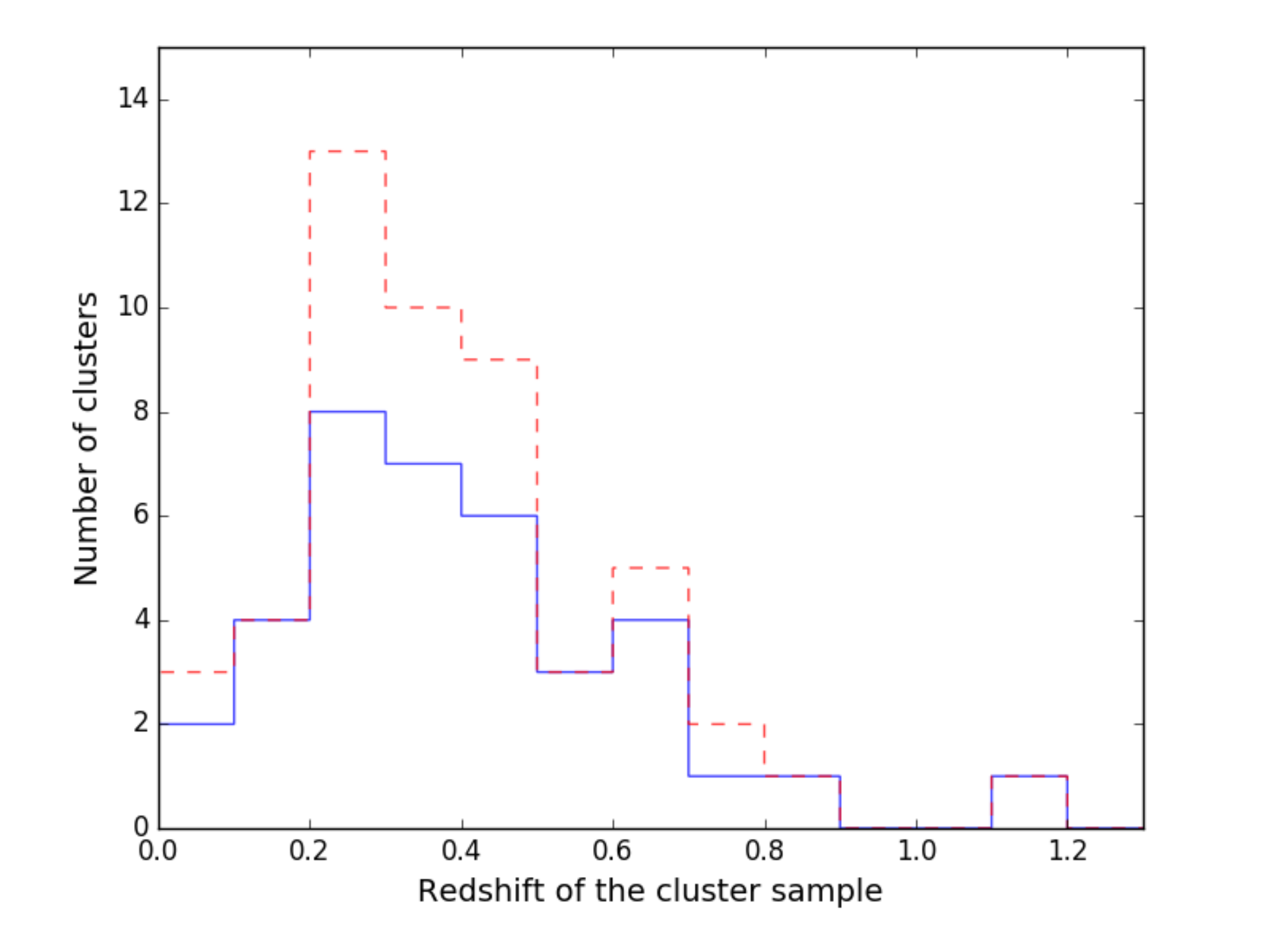}
\caption{Redshift distributions. Red dashed: full sample of 51 clusters with
  spectroscopic redshifts, blue solid: subsample of 37 clusters considered to
  investigate the ${\rm L_X-T_X}$ relation}
\label{fig:z_hist}
\end{figure}

The present study is based on our published cluster catalogue that comprises 54 galaxy groups/clusters in the redshift range from 0.05 to 1.2 with a median redshift of 0.36.The redshifts of these clusters were obtained from cross-correlated X-ray and optical cluster catalogues or from the NED.
We confirm published redshift values, based on photometric and spectroscopic data available in the SDSS. A spectroscopic confirmation based on at least one member galaxy with spectroscopic redshift is available for 51 clusters of our
sample. Fig.~\ref{fig:z_hist} shows the cluster redshift distribution
for the 51 clusters with spectroscopic redshifts and for the subsample
of  37 clusters with X-ray data of sufficient quality to allow the determination of the X-ray temperatures and luminosities that are used to investigate the ${\rm L_X-T_X}$ relation (see Sect.~3.2).

About two thirds of the cluster sample were known in previous X-ray selected cluster catalogues \citep[e.g.][]{Mehrtens12, Takey13, Takey14} while the remaining systems are newly discovered in X-rays. The X-ray  luminosities and masses of the clusters were estimated based on the fluxes given in the 3XMM-DR5 catalogue. The galaxy cluster catalogue is availble at the CDS\footnote{http://cdsarc.u-strasbg.fr/viz-bin/qcat?J/A+A/594/A32}. 

\section{X-ray properties of the cluster sample}

We present here  our procedure to reduce and analyse the XMM-Newton observations of the cluster sample. Since the X-ray data quality is not sufficient to determine the X-ray temperature profiles of the systems, we compute the global  temperatures and luminosities in a radius of 300 kpc. We expect to derive  X-ray temperatures and luminosities with reasonable uncertainties for about two thirds of the cluster sample that have more than 300 photon counts. These measurements will be used to investigate the X-ray luminosity-temperature (${\rm L_X-T_X}$) relation, as described below.     

\subsection{X-ray data reduction and analysis}

The 54 galaxy clusters constituting our sample were detected in 31 XMM-Newton observations. A few clusters  are detected in more than one XMM pointing. In this case, we choose the observation with the higher photon counts to extract the X-ray spectrum. The observation data files (ODFs, raw data) were downloaded using the Archive InterOperability System (AIO), which provides access to the XMM-Newton Science Archive (XSA). Both the data reduction and analysis of the sample were carried out using the XMM-Newton Science Analysis Software \citep[SAS:][]{Arviset02} version 15.0.0, following the recommended standard pipelines in the SAS manuals. To reduce the ODFs, we first generated the calibrated event list for the EPIC (MOS1, MOS2, PN) cameras using the latest calibration data. This step was done with the SAS tasks 
$\tt{cifbuild, odfingest, epchain, emchain}$. 

We then filtered the calibrated event lists by excluding observing intervals with high background flares and bad events. To do this, we followed 
the procedure recommended in the user guide of SAS, which has the following steps; (i)~We first created a light curve of the event file to check for bad pixels and columns, and high-background periods. (ii)~We then created a Good Time Interval (GTI) file that contains the good times corresponding to a background count rate that is approximately constant and low. This GTI file is used to filter the event list. (iii)~We applied the standard filter expression and the GTI to create a filtered event list. (iv)~Finally, we created a second light curve of the filtered event list to check the removal of high-background periods.  
The filtered calibrated event lists were used to create sky images in different energy bands. These last steps were done with the SAS packages $\tt{evselect, tabgtigen, xmmselect}$. 

The X-ray spectra of clusters were extracted from the EPIC filtered calibrated event lists within fixed circular apertures of radius 300 kpc centered on the X-ray emission peaks. This fixed source aperture was chosen because the spectral analysis could not be achieved with reasonable accuracy within $R_{500}$ for most of the cluster sample due to their X-ray data quality.   $R_{500}$ is the radius at which the cluster average density equals 500 times the critical density of the Universe estimated at the cluster redshift. 

A background spectrum for each cluster is also extracted in a fixed annulus with inner and outer radii equaling three (900 kpc) and four (1200 kpc) times the source extraction radius (300 kpc), respectively. Other sources overlapping the cluster circular and background annulus apertures are excluded from the regions used to extract spectra. The SAS meta task $\tt{especget}$ is used to generate the cluster and background spectra and to create the response matrix files (redistribution matrix file (RMF) and ancillary response file (ARF)) that are required to perform the X-ray spectral fitting. 

Before any fit, the photon counts of the cluster spectra are grouped into bins with at least one count per bin (as e.g. in \citealp{Takey13,Ogrean16})
using the Ftools task $\tt{grppha}$. For spectral fitting, we use the XSPEC version 12.9.0n \citep{Arnaud96} run by Python module $\tt{Pyspec}$. 

The EPIC spectra of each cluster are simultaneously fit by a combination of the $\tt{TBABS}$ absorption model \citep{Wilms00} and of a single-temperature optically thin thermal plasma $\tt{APEC}$ model \citep{Smith01}. In the fitting process, we fix the  Galactic hydrogen density column (nH) to the value derived from the Leiden/Argentine/Bonn (LAB) survey \citep{Kalberla05}. We also fix both the cluster redshift to the value given in the cluster catalogue and the metallicity to 0.3 $Z_\odot$. 
We used the spectroscopic redshifts for 51 systems and the photometric redshifts for the remaining three clusters.

The free parameters of the $\tt{APEC}$ model are the X-ray temperature and the spectral normalization. We use the Cash statistics in the fitting process and the energy range is [0.3-7] keV. We limited the energy range to 0.3-7 keV because the XMM telescope is poorly calibrated at energies softer than 0.3 keV and the cluster spectra are background dominated at energies higher than 7~keV \citep{Lloyd-Davies11}.
The results are the cluster X-ray temperature, aperture flux and luminosity (rest frame) in the [0.5-2] keV band, and their corresponding errors. The errors on the fit parameters are given in the 68 percent confidence range. 
We also derive the bolometric X-ray flux and luminosity (rest frame) in [0.1-50] keV from the dummy response matrices based on the best fit parameters. The bolometric flux and luminosity are derived with no errors. Here, we assume that the relative error on the bolometric luminosity is  the same as that on the luminosity in the [0.5-2] keV band produced by the fit, and the errors on the bolometric flux and luminosity are estimated in this way.
To make sure this assumption is valid and the resulting luminosity
does not depend too much on temperature errors, we varied the temperatures by $\pm 1\sigma$ in a few cases and found that the measured band luminosities are within their errors.


\subsection{The X-ray luminosity-temperature (${\rm L_X-T_X}$) relation}

\begin{figure}
\includegraphics[width=\columnwidth]{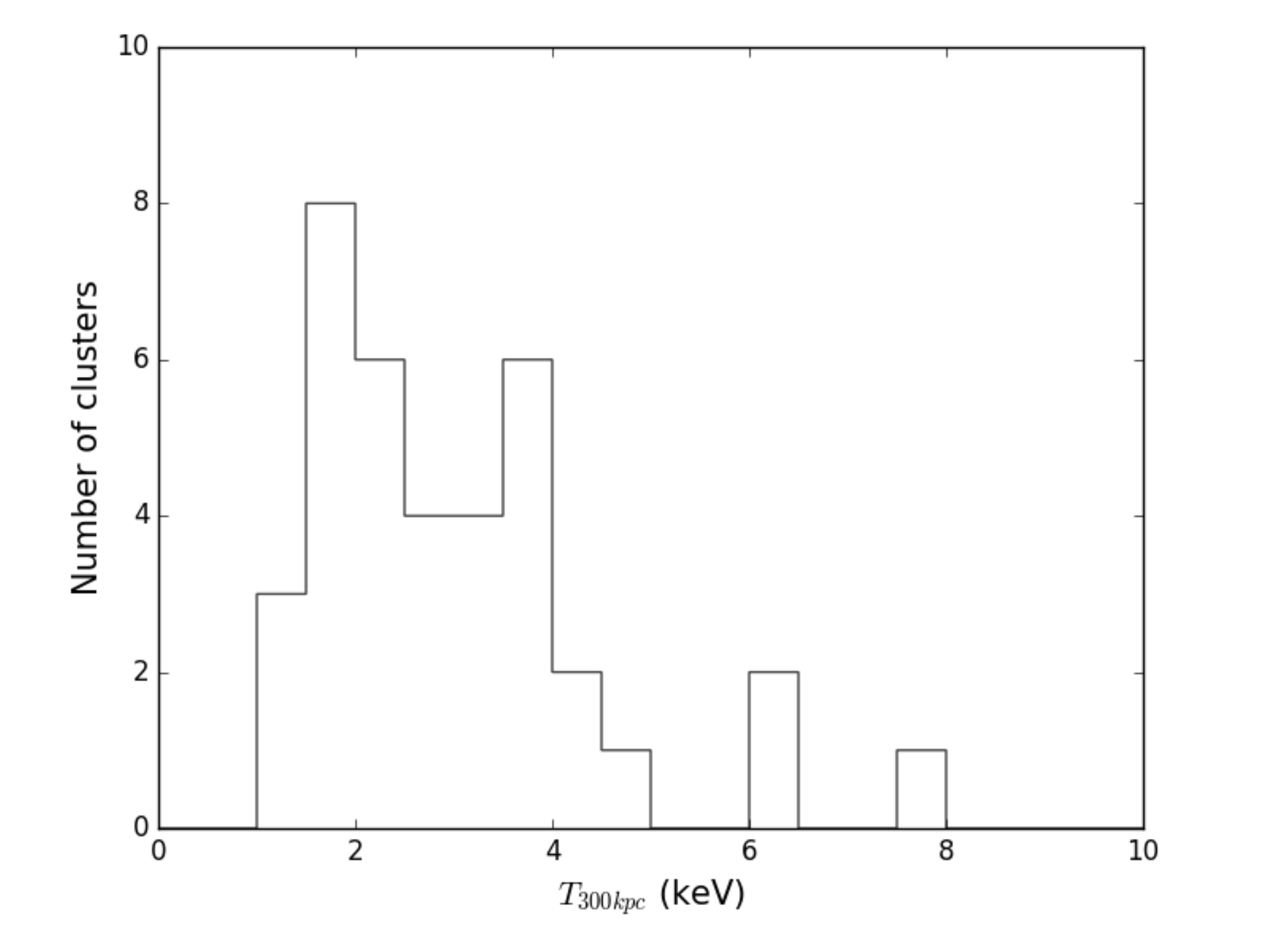}
\caption{Distribution of X-ray temperatures for 37 clusters used in ${\rm L_X-T_X}$ relation}
\label{fig:kT_hist}
\end{figure}

In the ${\rm L_X-T_X}$ relation, we only include the 37 galaxy clusters (69\%) from the cluster sample that have relative errors on temperatures and luminosities smaller than 50\%. This was done to obtain a relation with a slope and an intrinsic scatter unaffected by large uncertainties on temperatures and luminosities. The properties of the 37 clusters considered to compute the ${\rm L_X-T_X}$ relation are given in the Appendix (Table A1). The median, mean, and standard deviation of the temperature relative errors are  22\%,  21\%, and 11\%, respectively. The majority of the clusters in this   subsample have low temperatures, below 4 keV. Fig.~\ref{fig:kT_hist} shows the distribution of X-ray temperatures of the studied sample.

The redshift range of the cluster sample (37 systems) is from 0.07 to
1.2, with a median redshift of 0.36. There are 10 distant clusters in
the cluster sample with redshifts beyond 0.5. Figure~\ref{fig:z_hist}
shows the cluster redshift distribution for the systems used in the
${\rm L_X-T_X}$ relation. 

\begin{figure}
\includegraphics[width=\columnwidth]{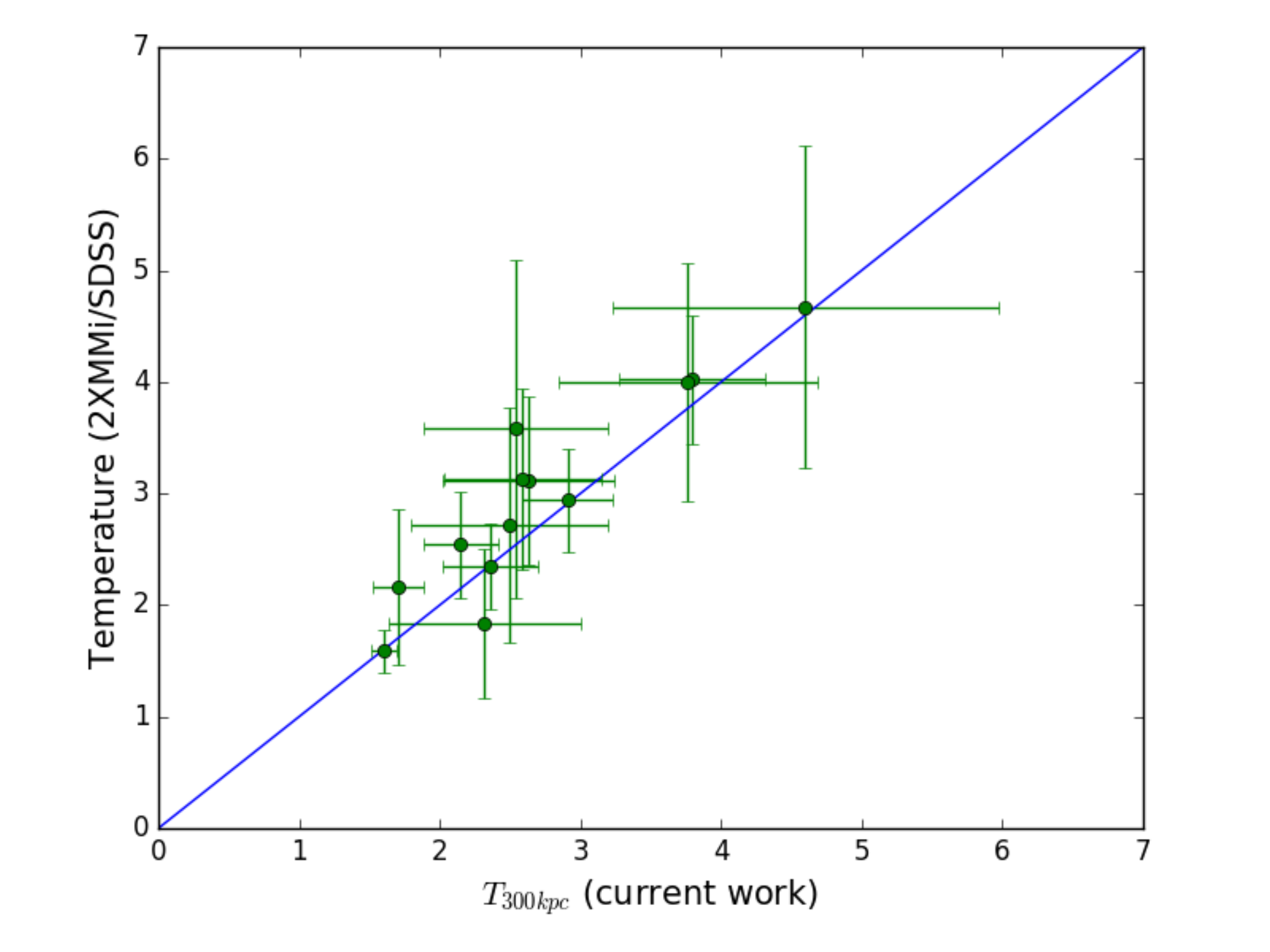}
\caption{Comparison of X-ray temperatures measured in our study at a radius of  300 kpc and in the 2XMMi/SDSS survey at aperture that maximizes the signal-to-noise ratio. The plotted errors are the average of the positive and negative errors provided by the spectral analysis. The solid line shows the one-to-one relationship.}
\label{fig:kT_comp}
\end{figure}

To check our results on cluster temperatures, we compare the temperatures derived within 300 kpc with those published within a different aperture (that maximize the signal-to-noise ratio) in the 2XMMi/SDSS catalogue  by \citet{Takey13}. Figure~\ref{fig:kT_comp} shows a good agreement with no systematics for the 13 clusters in common between our cluster sample and the 2XMMi/SDSS cluster sample. 

The advantage of having derived temperatures within an aperture of 300 kpc, is that it allows a direct comparison of our ${\rm L_X-T_X}$ relation with that published by \citet{Giles16}, who also determined the temperature within 300~kpc and the luminosity within $R_{500}$ for a sample of clusters of comparable redshifts. \citet{Giles16} investigated the ${\rm L_X-T_X}$  relation for the 100 brightest galaxy clusters detected in the XXL survey made by the XMM-Newton mission. Also, the temperatures within 300 kpc are comparable to the temperatures within apertures that represent the highest signal-to-noise ratio published by \citet{Takey13}, see Fig.~\ref{fig:kT_comp}. 

The X-ray temperature measurements within $R_{500}$ and 300~kpc are comparable, with no systematic differences, as shown by \citet{Giles16}. To check
if this agreement is valid in our cluster sample, we extracted spectra within $R_{500}$ for 15 systems with fluxes in [0.5-2] keV band higher than $5\times 10^{-14}$~erg~cm$^{-2}$~s$^{-1}$. The fluxes and $R_{500}$ values are obtained from the catalogue published by \citet{Takey16}. We then fitted the spectra with the same procedure as used in the current analysis. The ratio of the temperatures   within $R_{300~kpc}$ and $R_{500}$ has a mean and standard deviation of 1.035 and 0.258, respectively. This means that the temperatures  within $R_{300~kpc}$  are comparable to those  within $R_{500}$, since their mean increases by only 4\%, which is much smaller than the mean relative error on the temperature (20\%) for these 15 systems. 

To investigate the ${\rm L_X-T_X}$ relation between the bolometric luminosity within $R_{500}$ ($L_{500}$ hereafter) and the temperature within 300~kpc, we first need to determine $L_{500}$ based on the aperture bolometric luminosity within 300~kpc ($L_{300~kpc}$ hereafter).
We prefer to re-determine $L_{500}$ based on spectral fitting parameters from the present work, rather than taking $L_{500}$ from our earlier work \citep{Takey16}, that was based on the fluxes from the 3XMM-DR5 catalogue. 

The computing of $L_{500}$ is done through an iterative method to extrapolate the aperture (300 kpc) bolometric flux to the $R_{500}$ bolometric flux, $F_{500}$, by appplying the beta model, a hydrostatic isothermal model used to describe the X-ray surface brightness profiles $S(r)$ of galaxy clusters:
\begin{equation}
S(r) = S(0)\ \Bigg[1 + \Big(\frac{r}{r_{\rm c}} \Big)^2 \Bigg]^{-3\beta + 1/2},
\end{equation}
where r$_{\rm c}$ is the core radius. The model assumes that both the hot intracluster gas and the cluster galaxies are in hydrostatic equilibrium and isothermal.

To do this, we first computed the cluster mass within $R_{500}$, $M_{500}$, based on the $L_{500} - M_{500}$ relation from \cite{Pratt09}. The first input for this relation is the aperture bolometric luminosity, $L_{300~kpc}$. The output $M_{500}$ is used to compute a first estimate of $R_{500}$. The $L_{300~kpc}$ luminosity is also utilized to compute the cluster temperature at $R_{500}$, $T_{500}$, based on the $L_{500} - T_{500}$ relation from \citet{Pratt09}. The estimated value of $T_{500}$ is then considered to compute the cluster core radius and beta value based on published relations by \citet{Finoguenov07}. The beta model is then applied to calculate the fluxes enclosed within the aperture and within $R_{500}$. The ratio of the aperture to $R_{500}$ fluxes is utilized to extrapolate $L_{300~kpc}$ to the $R_{500}$ luminosity. This extrapolated luminosity is then considered as an input for another iteration and all computed parameters are updated. This iterative procedure is repeated until converging to a final solution, where the flux within the new estimated $R_{500}$ is the same as the previous flux in the iteration. At this stage, we computed the bolometric luminosity, $L_{500}$, that is used in investigating the ${\rm L_X-T_X}$ relation. The output luminosities, $L_{500}$, derived by this iterative method are comparable to the ones determined in the XMM Cluster Survey (XCS) by \citet{Mehrtens12}. The details of this method, the scaling relations, and the comparison of $L_{500}$ were described by \citet[][]{Takey11, Takey13}.

We fit the ${\rm L_X-T_X}$ relation for our cluster sample using the BCES orthogonal regression method \citep[bces Python module, ][]{Akritas96} taking into account the errors on the luminosity and temperature as well as the intrinsic scatter of the relation. It is important to take into account the intrinsic scatter/dispersion  of the ${\rm L_X-T_X}$ relation since the data points do not lie exactly on a straight line and this line is not of slope 1. 
The fit is applied to the 37 clusters with relative errors on the temperatures and luminosities smaller than 50\%. 

\begin{figure}
\includegraphics[width=\columnwidth]{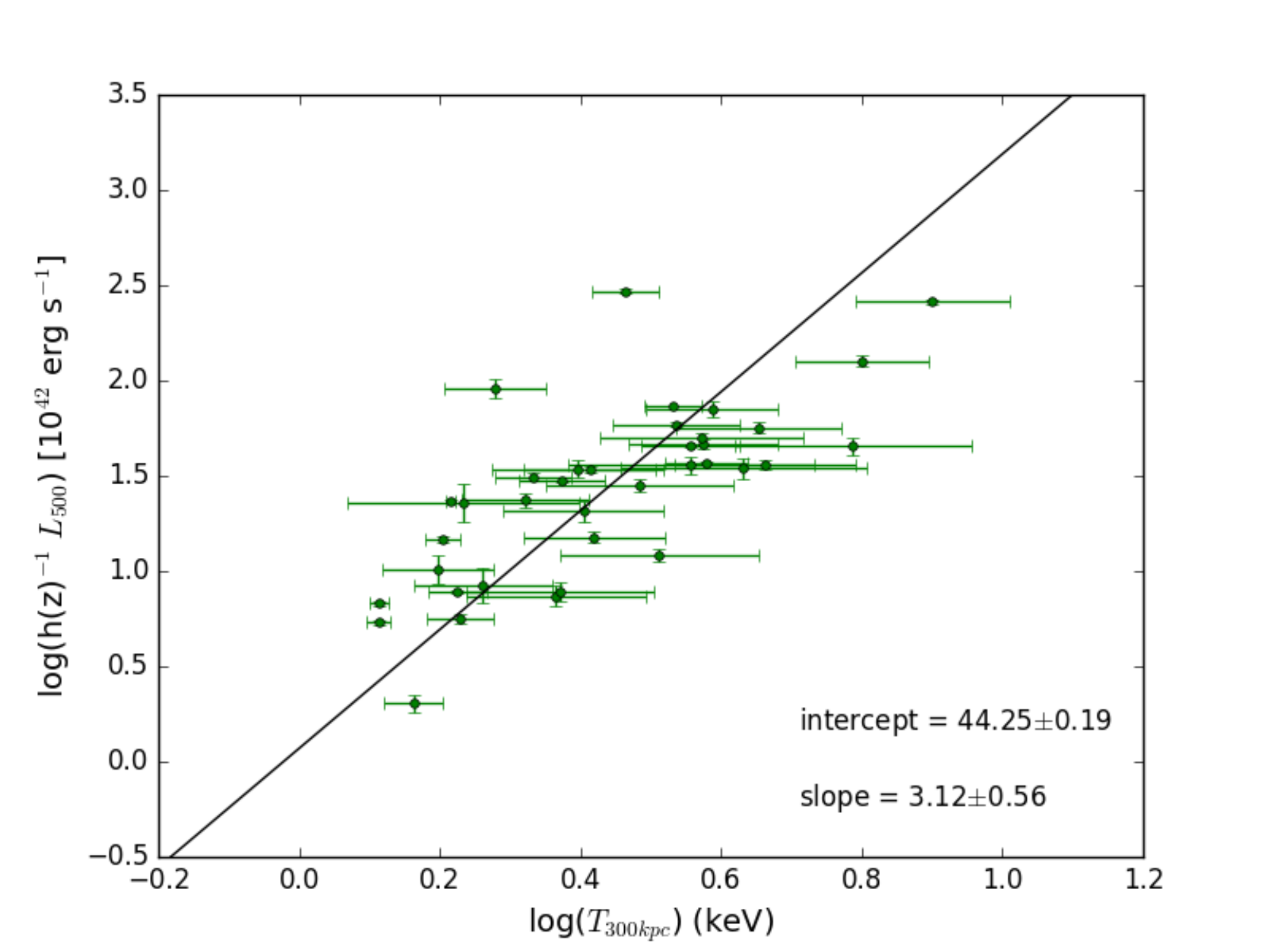}
\caption{X-ray bolometric luminosity within $R_{500}$, $L_{500}$, plotted against X-ray temperature within 300 kpc for the 37 galaxy clusters that have relative errors on their luminosity and temperature smaller than 50\%. The solid line represents the best fit to the data using a python module of the BCES orthogonal regression. The slope and intercept are written in the lower right corner. } 
\label{fig:3XMM_L500hz_T}
\end{figure}

Figure~\ref{fig:3XMM_L500hz_T} shows the ${\rm L_X-T_X}$ relation for our cluster sample. The best fit slope ($3.12\pm0.56$) is in agreement with the value ($3.03\pm0.28$) derived for the 100 brightest clusters in the XXL project published by \citet{Giles16}, but our slope has a larger uncertainty, possibly due to the X-ray data quality.
In addition, the quality of the data did not allow us to exclude the cluster core when extracting the spectra. \citet{Pratt09} showed that the scatter in the relation is reduced by more than a factor of two when excluding the cluster central regions.    
We also find good agreement with the ${\rm L_X-T_X}$ relation slopes in the literature \citep[e.g. ][]{Pratt09, Mittal11, Hilton12, Takey13, Rabitz17}.

Table~\ref{tab:LTcomp} shows the slopes and intercepts of the ${\rm L_X-T_X}$ relations evaluated for various cluster samples in different redshift ranges including our cluster sample. It can be noticed that the slope from the present work agrees within one sigma error with the published ones.
Regarding the intercept of the ${\rm L_X-T_X}$ relation, we find that the current value ($44.25\pm0.19$) is in agreement with those published by \citet[][]{Pratt09, Takey13, Giles16} within one sigma and within two sigmas with those by \citet{Mittal11, Hilton12}.

It is also noticed from the ${\rm L_X-T_X}$ relation (Figure~\ref{fig:3XMM_L500hz_T}) that the data points are scattered around the fitted line. To determine the intrinsic scatter in the luminosity
$\sigma_{\rm log}L_{500}$, we followed the procedure utilized by \citet{Pratt09}. In that method, the raw scatter is first determined using the error-weighted orthogonal distances to the regression line \citep[see Equations (3) and (4) in][]{Pratt09}. Then the intrinsic scatter of the luminosity is computed as the mean value of the quadratic differences between the raw scatters and the statistical errors of luminosity. The intrinsic scatter error is determined as the standard error of its value. This yields that the intrinsic scatter of the luminosity
$\sigma_{\rm log}L_{500}$ in the current ${\rm L_X-T_X}$ relation is $0.54\pm0.09$, which is higher than the value ($0.32\pm0.06$) of the REXCESS sample \citep{Pratt09} and the one ($0.48\pm0.03$) of the 2XMM/SDSS sample \citep{Takey13}. In a similar way, we compute the intrinsic scatter of temperature $\sigma_{\rm log}T_{300~kpc}$ ($0.14\pm0.02)$, which is also higher than the one ($0.07\pm0.01$) for the HIFLUGCS sample derived by \citet{Mittal11}.

We also derived the slope and intercept of the ${\rm L_X-T_X}$ relation in clusters of low ($z<0.3$) and high ($z\geq 0.3$) redshift and found values similar to those for the whole sample, within the error bars. This agrees with the fact that the  slopes and intercepts found in the literature for various redshift ranges are comparable (see Table~\ref{tab:LTcomp}).

As mentioned above, our cluster survey is based on 94 X-ray cluster candidates selected from the 3XMM-DR5 extended sources that are located in the SDSS S82 region. Since the 3XMM catalogue is based on XMM observations with a wide range of exposure times, it is not an easy task to assess the completeness of the list of extended sources in this catalogue or to assess the selection function. The catalogue may miss some extended sources with low photon counts or large core radii, or may include them with incorrect parameters. This implies that our X-ray cluster candidate list is not a complete one, and that it is not a flux-limited sample. The effect of the selection function on the ${\rm L_X-T_X}$ relation cannot therefore be estimated from the current sample. Thus, checking the evolution of the relation is not possible. 

Also, only 54 systems have been optically confirmed with redshift estimates. Of these, 37 clusters have a sufficient data quality to investigate the ${\rm L_X-T_X}$ relation. Therefore, there are missing clusters with measured redshifts ($54-37=17$ systems) and missing candidates with no redshift estimate ($94-54=40$ candidates).   
The majority of the missing clusters and/or candidates in the relation are distant objects that may have no significant effect on the slope of the relation \citep{Hilton12}. However, if the missing clusters and/or candidates include galaxy groups with low luminosities and temperatures, this can make the slope of the relation shallower \citep{Takey13}.

\begin{table}
\caption{ Comparison of the intercept and  slope of the ${\rm L_X-T_X}$ relation with those published in the literature (N$_{ClGs}$ is the number of clusters considered in the relation.) }

\label{tab:LTcomp}
\begin{tabular}{c r c c c}
\hline\hline
 Redshift   & N$_{ClGs}$& Intercept      & Slope         & References\\
 range      &           &                &               &     \\
\hline
0.07 - 1.20 & 37        & $44.25\pm0.19$ & $3.12\pm0.56$ & 1\\
0.04 - 1.05 & 100       & $44.10\pm0.03$ & $3.03\pm0.28$ & 2\\
0.06 - 0.25 & 96        & $44.63\pm0.10$ & $3.18\pm0.22$ & 3\\
0.004 - 0.22& 64        & $44.70\pm0.03$ & $2.94\pm0.16$ & 4\\ 
0.06 - 0.18 & 31        & $44.85\pm0.70$ & $3.35\pm0.32$ & 5\\
0.03 - 0.67 & 345       & $44.39\pm0.06$ & $2.80\pm0.12$ & 6\\ 
\hline
\end{tabular}
References: 1. Present work; 2. \citet{Giles16}; 3. \citet{Hilton12}; 4. \citet{Mittal11}; 5. \citet{Pratt09}; 6. \citet{Takey13}.
\end{table}



\section{Morphological analysis of cluster galaxies}

To study the morphological properties of the galaxies belonging to the
clusters of our sample, we also limited our analysis to the 54
clusters with measured redshifts (51 spectroscopic and 3
photometric). Details are given in the next subsections, but we
briefly summarize our method here. First, we extract the images
covering each cluster, model the PSF and measure for each galaxy the
flux in the bulge and in the disk, to classify each galaxy as early
type or late type. The detected objects are matched with existing
spectroscopic and photometric redshift catalogues.  Second, we extract
the galaxies within two circular zones around each cluster: a large
region of 2~Mpc radius, and a smaller region within $R_{200}$. 
The latter quantity is estimated with the relation 
$R_{200}=1.5\times R_{500}$ as derived from clusters in the XMM cluster survey data release one \citep{Mehrtens12}, with $R_{500}$ obtained from the galaxy cluster catalogue published by \citet{Takey16}. The values of $R_{200}$ are
given in Table C1. 
Third, we select in these two regions
the galaxies with a high or relatively high probability of belonging
to the cluster, according to their spectroscopic or photometric
redshifts, respectively (see 4.1.2). These galaxies are used to
compute the fraction of early and late type galaxies as a function of
cluster mass and distance to the cluster X-ray centre by stacking the
clusters respectively in mass bins and in radial bins.


\subsection{The method}

\subsubsection{Extraction of cluster images and galaxy measurements}

We extract the images from the IAC Stripe 82 Legacy Project conducted by \citet{Fliri16}\footnote{available at
  http://www.iac.es/proyecto/stripe82/index.php} in the five bands u',
g', r', i' and z', as well as in a band called rdeep which is the sum
of every observation in the g', r' and i' bands. The latter band is
not photometrically calibrated, but we retrieve it to detect and
characterize faint objects. Each image covers $0.25 \times
0.25$~deg$^2$ with a pixel size of 0.396~arcsec. Since most clusters
do not fall at the center of one image, we assemble 4 images
per cluster and per filter. For the three clusters with the smallest
redshifts, we assemble 9 images in order to cover a circle of
2~Mpc radius at the cluster redshift.  The images are assembled with
the SCAMP and SWARP softwares developed by \citet{Bertin10}\footnote{available at http://www.astromatic.net/}.  The
photometric zero points are calculated by applying Eq.~(7) from \citet{Fliri16}.

The images in the five bands are used to derive the galaxy luminosity
functions presented in Section 5.  For the morphological study
presented in this section, we limit our analysis to the r' band to
save computing time. This is justified by the fact that in our
previous paper \citep{Durret15} we found that the results in the g'
and i' bands were very similar to those in the r' band.  We did not
attempt to use the rdeep images, because since they are the sum of
images in three bands their PSF is not as accurate as for a single
band, and besides they are not calibrated photometrically.

All the objects are detected on each image with SExtractor
\citep{Bertin96}.  We then run PSFEx \citep{Bertin11}, a software that
takes as input a catalogue of objects detected by SExtractor and
models the Point Spread Function (PSF). By injecting the PSF models
into SExtractor again and comparing them to the original image, the
program fits 2D photometric models to the detected objects.  We
eliminate stars by keeping only the objects with the SExtractor
parameter CLASS\_STAR$<0.95$. The fitting process is very similar to
that of the GalFit package \citep{Peng02} and is based on a modified
Levenberg-Marquardt minimization algorithm. The model is convolved
with a supersampled model of the local point spread function (PSF),
and downsampled to the final image resolution. The PSF variations are
fit using a six--degree polynomial of $x$ and $y$ image
coordinates. In this way, we obtain for each galaxy the fluxes in the
bulge (a de Vaucouleurs spheroid) and in the exponential disk.

We consider S\'ersic surface brightness models with two components, a de
Vaucouleurs bulge:
\begin{equation}
\Sigma = \Sigma_e \ \mathrm{exp}\left (-7.67 \left [\left (\frac{r}{r_e} \right )^{1/4}-1 \right ] \right ) 
\end{equation}
and an exponential disk:
\begin{equation}
\Sigma = \Sigma_0 \ \exp\left (\frac{r}{r_d} \right ). 
\end{equation}

We thus obtain a catalogue of relatively bright objects, containing
for each galaxy its coordinates, flux in the disk $f_{disk}$ and flux
in the bulge $f_{bulge}$, and magnitude (MAG\_MODEL), computed by
  SExtractor from the sum of the disk and bulge fluxes. Since our
final goal is to separate early and late type galaxies, we only keep
the galaxies with a relative error on the model fluxes smaller than
15\% (as in \citealp{Durret15}). The flux ratio of the two components
allows a classification into early and late type galaxies: early type
galaxies are those with
$f_{spheroid}/(f_{disk}+f_{spheroid}) \geq 0.35)$ and late types are
those with $f_{spheroid}/(f_{disk}+f_{spheroid}) < 0.35)$, as in
\citet{Simard09}.


\subsubsection{The final galaxy catalogue}

To select the galaxies with a high probability of belonging to each of
the 54 clusters with redshifts available in our sample, we must
assign a redshift to each galaxy of the morphological catalogue. This
is done in two steps, because two different catalogues were
available: the SDSS DR12 catalogue includes spectroscopic redshifts
for some galaxies and photometric redshifts for many relatively bright
galaxies, while the \citet{Reis12} catalogue gives better quality
photometric redshifts, but only for objects fainter than $r=16$, and
goes deeper than DR12.

When a spectroscopic redshift is available, we assign it to the
corresponding galaxy.  If not, we assign the DR12 photometric
redshift to galaxies with $r<16$ and the \citet{Reis12} photometric 
redshift to galaxies with $r\geq 16$.
This allows us to obtain a large catalogue containing for each galaxy:
the coordinates, spectroscopic (when available) or photometric
redshift, the r' band magnitude, the flux in the disk, the flux in the
bulge, and the uncertainties on those parameters.

For each cluster, we then select from this catalogue the galaxies
within two different radii from the X-ray center, in projection on
the plane of the sky, within a circular aperture: a large radius of
2~Mpc and within a smaller radius of $R_{200}$.  
Finally, for each cluster, we apply a selection criterion of cluster membership based on
the redshift: we only keep galaxies with spectroscopic redshifts
differing from the cluster redshift, $z_{cluster}$, by less than
$\pm 0.01$, and galaxies with photometric redshifts differing from
that of the cluster by less than $\pm 0.03(1+z_{cluster})$, as in
\citet{Takey16}.

Therefore, for every cluster, we obtain two catalogues of cluster
galaxies, one within 2~Mpc and one within $R_{200}$.  The latter
catalogue is a subset of the former and will be used to compute
galaxy luminosity functions in Section~5. 


\subsubsection{Selection of the brightest cluster galaxy}

Since the position of the X-ray emission peak is known for each
cluster, we have identified the brightest cluster galaxy (BCG) as the
brightest galaxy (or one of the brightest galaxies) located the
closest to the X-ray peak. For relaxed clusters, the BCG is expected
to be located very close to the X-ray centre. However, a few clusters of
our sample are very close to each other and at comparable
redshifts. In this case, they may be merging and the BCG may be
displaced from the X-ray maximum, so determining which galaxy is the
BCG can be more difficult. For the sake of completeness, we have
identified the BCGs of the individual clusters and we list them
in Table~B1. 


\subsection{Results}

We compute the fraction of early and late type galaxies as a function
of cluster mass and distance to the cluster X-ray centre. For
this we stack the clusters respectively in mass bins and in radial
bins. Our results are given below.

\begin{figure}
\includegraphics[width=\columnwidth]{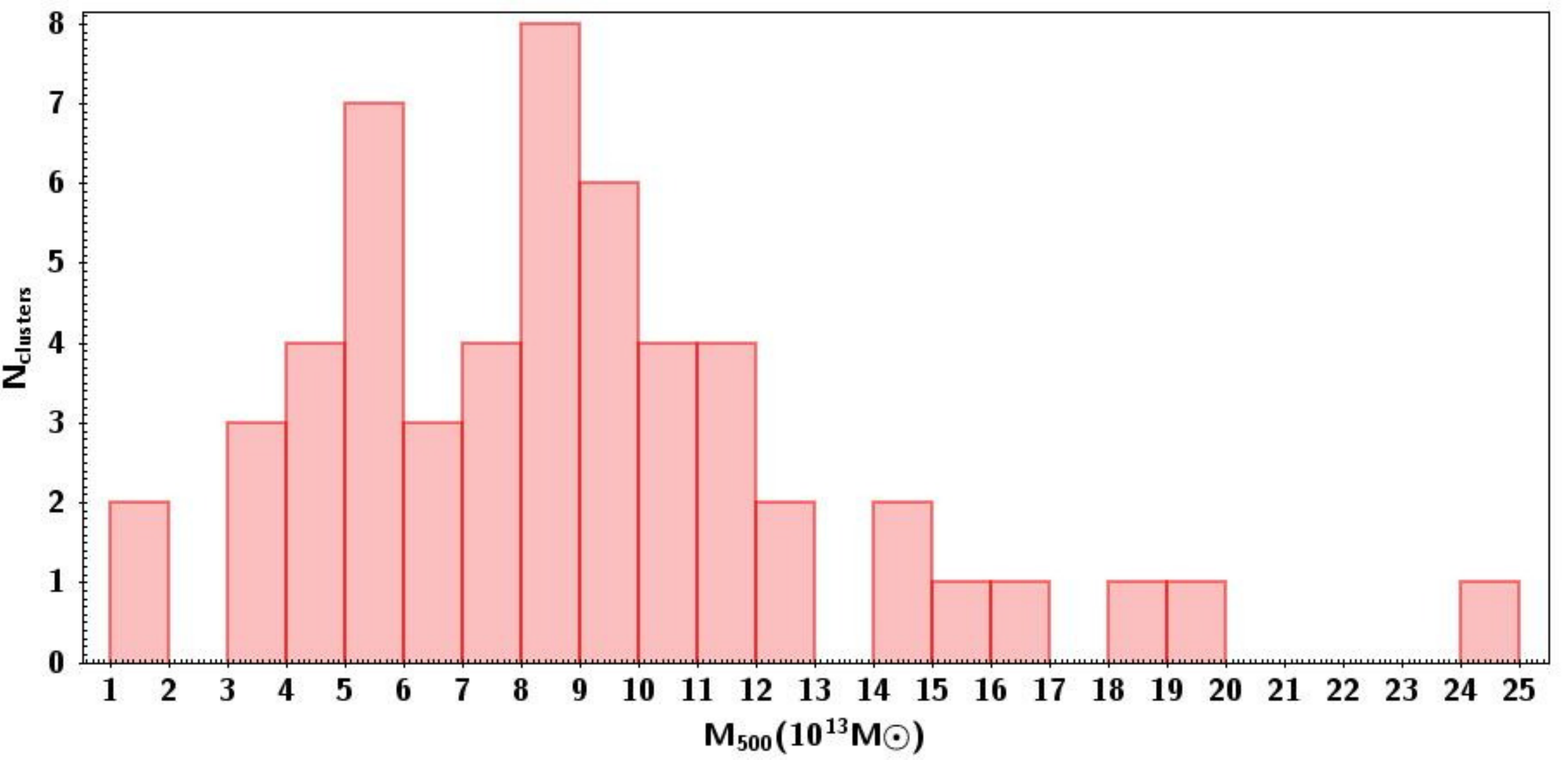}
\caption{Histogram of the cluster masses within $R_{500}$, $M_{500}$, in units 
of $10^{13}$~M$_\odot$ computed from the X-ray data.  }
\label{fig:histo_mass}
\end{figure}

\begin{figure}
\resizebox{\hsize}{!}{\includegraphics[viewport= 25 0 680 500, clip]{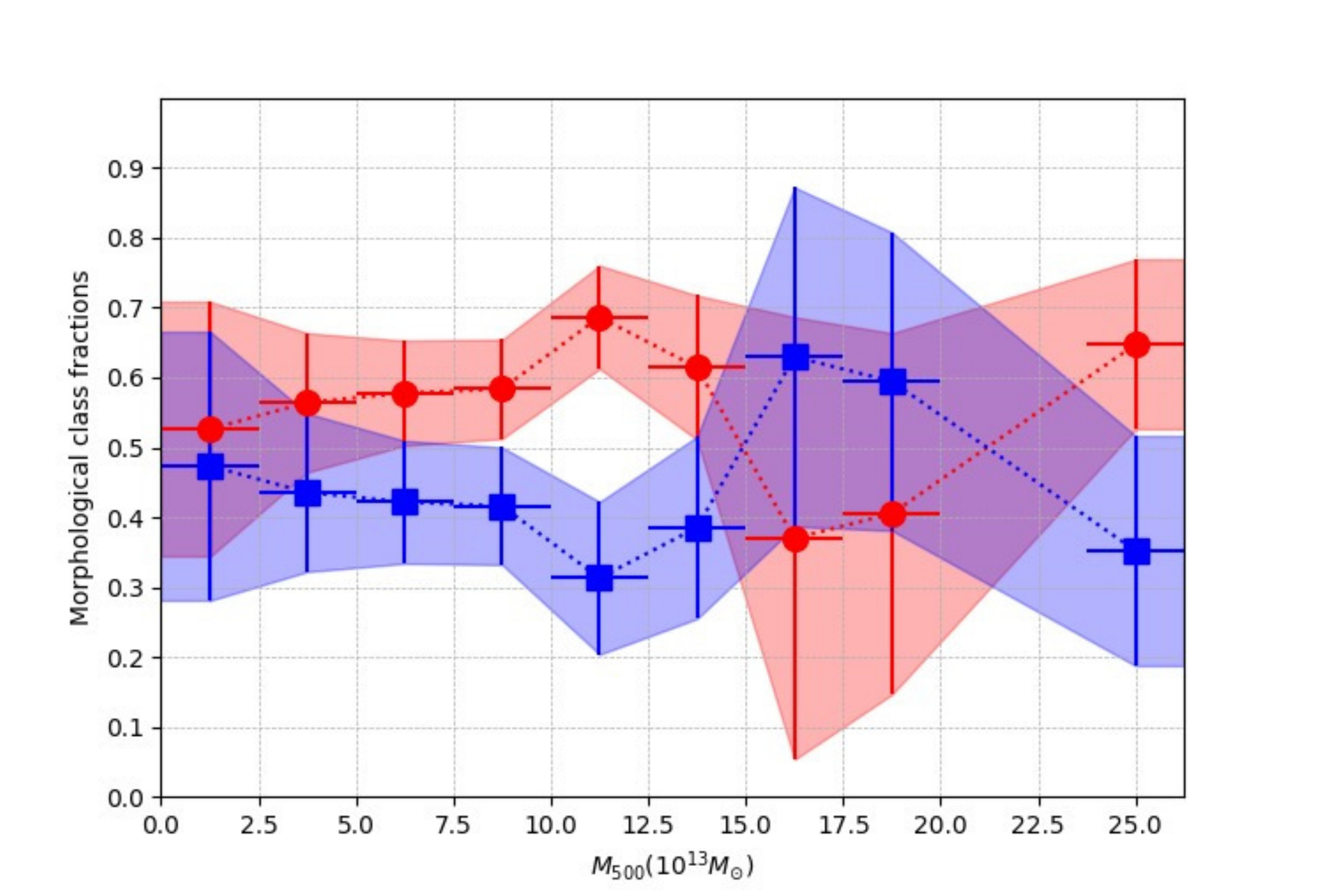}}
\caption{Fraction of early-type (red circles) and late-type (blue squares) as a 
function of cluster mass obtained after stacking the 54 clusters in mass bins.}
\label{fig:fractype_mass}
\end{figure}

\begin{figure}
\resizebox{\hsize}{!}{\includegraphics[viewport= 15 0 600 500, clip]{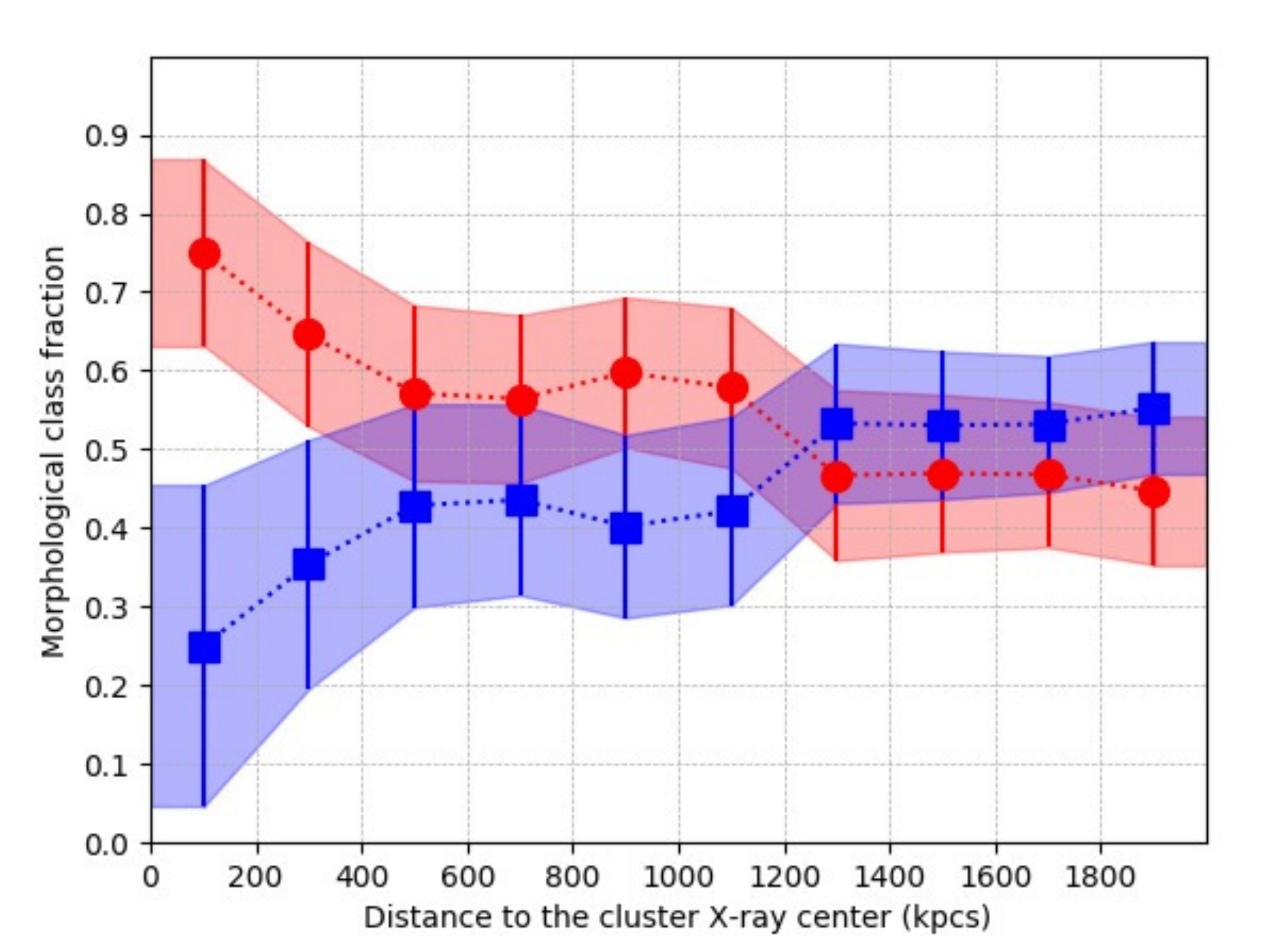}}
\caption{Fraction of early-type (red circles) and late-type (blue squares) as a function of distance to the cluster center 
  obtained after stacking the 54 clusters in ten bins. }
\label{fig:fractype_r}
\end{figure}

The histogram of the cluster masses within $R_{500}$, $M_{500}$, is shown
in Fig.~\ref{fig:histo_mass}.  We compute the fractions of early and
late-type galaxies in ten $M_{500}$ mass bins
and show the corresponding results in
Fig.~\ref{fig:fractype_mass}. For each bin in cluster mass (and later
distance to the cluster centre) in Fig.~\ref{fig:fractype_mass} (and
later Fig.~\ref{fig:fractype_r}), the error bars are calculated
considering Poisson distributions, hence as $\sqrt{N}/N$, where $N$ is
the number of galaxies in each bin. No strong variation is observed,
except perhaps for the most massive clusters, where there seems
  to be a somewhat larger fraction of late-type galaxies in the range
  $15\times 10^{13} < {\rm M}_{500} <20\times 10^{13}$~M$_\odot$.
  However, since this is not the case in the bin corresponding to the
  highest mass, it is difficult to say if there is a general trend and
  to give an interpretation.

The fractions of early and late type galaxies were also computed as a
function of cluster radius (in ten bins). 
The results are shown in Fig.~\ref{fig:fractype_r}. As
  expected, the fraction of early types is very large (close to 80\%)
  in the innermost bins and decreases down to $\sim$50\%  
  around 1.3~Mpc, while the fraction of late types increases with
  radius and becomes larger than 50\% around 1.3~Mpc.

A certain amount of contamination by foreground and background
  galaxies must occur when considering the fractions of early and late
  type galaxies. We estimated this contamination by comparing the red
  sequence galaxy counts in each magnitude bin and the corresponding
  background counts from the COSMOS survey, as estimated in Section~5
  to compute GLFs. Values of the contamination vary from one cluster
  to another between 30\% and 70\% at a magnitude of $r'\sim 20$ with
  no obvious dependence on redshift or on the M$_{500}$ cluster mass.
   The signal dilution due to contamination  is expected to be
    stronger for low mass clusters, for which the contrast above the
    field is lower. This could explain our finding that low mass poor
    systems (Fig.~\ref{fig:fractype_mass}) or cluster outermost parts
    (Fig.~\ref{fig:fractype_r}) have early to late type fractions
    comparable to those of the field. 

We also tried to analyse the variations of the fractions of early and
late types as a function of the number of galaxies within $R_{200}$
but found no significant result. Neither did we find any significant
variation of the fractions of early and late types with redshift.



\section{The galaxy luminosity functions of cluster galaxies}

\subsection{The method}

\begin{figure} 
\resizebox{8 cm}{!}{\includegraphics{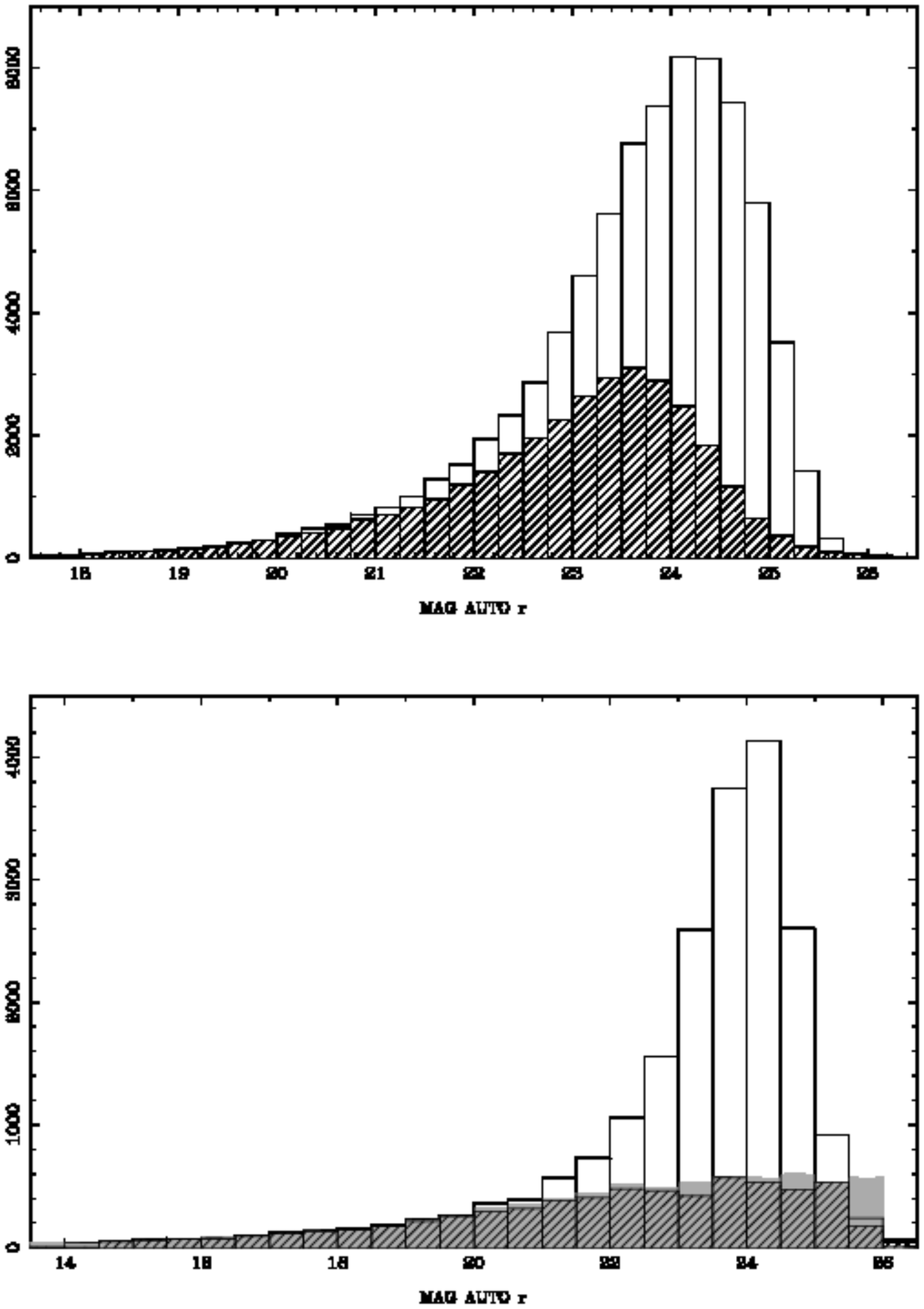}}
\caption{{\it Top:} galaxy magnitude histogram in the r band from
the IAC catalogue (hatched) and from our catalogue (white) for cluster 
3XMM~J001737.3-005240. 
{\it Bottom:} star magnitude histogram in the r band from
the IAC catalogue (white), from our catalogue (hatched) and from the
Besan\c con model counts (grey) for cluster 3XMM~J001737.3-005240.}
\label{fig:histomag_compar}
\end{figure}

We derive the galaxy luminosity functions (GLFs) of the 51 clusters
with spectroscopic redshifts from our sample. We first test the
quality of the \citet{Fliri16} catalogues.  For this, we retrieve for
one cluster (3XMM~J001737.3$-$005240) the galaxy catalogue by
\citet{Fliri16} in the cluster area and compare it to the one we
obtain with our own method, where we optimize the extraction
parameters (see description below).  The result is that with our
method we detect more faint galaxies above $r\sim 21$. This is
illustrated by Fig.~\ref{fig:histomag_compar}.  The top figure shows
the galaxy magnitude histogram in the r band from the IAC catalogue
and from our catalogue extracted as described below for cluster
3XMM~J001737.3-005240. We can see that above $r\sim 21$ we start
detecting more galaxies. This seems due to a difference in
galaxy--star separation between the two methods.  As a comparison, we
plot in the bottom figure the star magnitude histogram in the r band
from the IAC catalogue, from our catalogue and from the Besan\c con
model counts \citep{Robin03} for cluster 3XMM~J001737.3$-$005240. We
can see that our star counts match well those of the Besan\c con
model, while the star counts from \citet{Fliri16} are much higher at
faint magnitudes. Since our aim here is to go as deep as possible to
measure the faint end slope of the galaxy luminosity functions, we
decide to reextract catalogues from the images that we had already
retrieved for the morphological analysis (see previous Section). This
should be considered as a {\it caveat} to future users of the
\citet{Fliri16} catalogues.

As described in the previous section, the images are retrieved in the
five SDSS bands, plus rdeep. We make detections with SExtractor
\citep{Bertin96} in the rdeep band, then measure magnitudes
(MAG\_AUTO) in dual image mode, using rdeep as a reference.  The
photometric zero points are calculated by applying Eq.~(7) from
\citet{Fliri16}.  For some clusters, it was necessary to mask some
areas covered by very bright stars (and even one bright foreground
galaxy).  We then separate stars from galaxies based on a maximum
surface brightness versus magnitude diagram \citep{Jones91}. We
always check that the histogram of the number of objects classified as
stars is consistent with the number of stars predicted in the cluster
direction by the Besan\c con model quoted above.

For each cluster, we apply the following steps. We limit
our analysis to galaxies within the $R_{200}$ radius of each cluster,
for two reasons. First, this value is chosen to increase the contrast
over the background, and second it allows to separate better clusters
that are close in projection on the sky. The red sequence (RS) is
defined based on a colour-magnitude diagram. In order to bracket the
4000~\AA\ break, we choose different colour-magnitude diagrams for
different cluster redshift ranges: $g-r$ versus $r$ for $0<z<0.43$,
$r-i$ versus $i$ for $0.43\leq z \leq 0.70$ and $i-z$ versus $z$ for
$z>0.70$ \citep{Hao10}. Galaxies with spectroscopic redshifts within 0.01
of the cluster redshift
and with photometric redshifts within $\pm
0.04(1+z_{cluster})$ are superimposed on the colour-magnitude
diagrams to define better the RS.  The slope of the colour-magnitude
relation is fixed to $-0.0436$ \citep[as e.g. in][]{Martinet15}.  A
first estimation is made by eye. We then select all galaxies within $\pm
0.6$ of this fit to compute the best fit to the colour-magnitude
relation, and we keep all the galaxies within $\pm 0.3$ of this
best fit \citep[see e.g. in][]{De-Lucia07} to compute the GLF.

The subtraction of the background galaxy contribution is made using
the COSMOS catalogue of \citet{Laigle16}, which covers a region of
1.38~deg$^2$, more than 10 times larger than the zones covered by our
clusters. Magnitudes from the COSMOS catalogue (Subaru filters) are
transformed into SDSS magnitudes with LePhare
\citep{Arnouts99,Ilbert06}, with extinction laws by \citet{Calzetti99}
and emission lines from \citet{Polletta06}. We then extract for each
cluster the COSMOS background counts corresponding to the same
extraction around the RS as for cluster galaxies, normalize all the
counts to 1~deg$^2$ and make count histograms in bins of
0.5~magnitudes. This is done in all five bands, u, g, r, i and z.

\begin{table}
  \caption{90\% and 80\% completeness limits for the detections of extended sources in the five bands considered.  The last line gives the completeness limits of the SDSS Stripe~82 data given by \citet{Annis14}.}
\begin{tabular}{cccccc}
\hline
\hline
filter & u & g & r & i & z \\
\hline
90\%       & 22.9 & 23.5 & 23.1 & 22.6 & 21.7 \\
80\%       & 23.1 & 23.8 & 23.6 & 22.9 & 22.0 \\
\hline
Annis 90\% & 23.1 & 22.8 & 22.4 & 22.1 & 20.4 \\
\hline
\end{tabular}
\label{tab:completeness}
\end{table}

Before analysing GLFs, we estimate the completeness levels reached in
each band. This is done through point source simulations as in
\citet{Martinet15}. The completeness limits for extended sources are
about 0.5 magnitudes brighter than for point sources
\citep{Adami07}. We compute GLFs within the 90\% and 80\% completeness
limits given in Table~\ref{tab:completeness}.  We also give in
  this Table the 90\% completeness limits of the SDSS Stripe~82 given
  by \cite{Annis14}.  We can note that except in the u band the data
  extracted from \citet{Fliri16} appear deeper, thus justifying our
  choice.

Finally, apparent magnitudes $m$ are converted to absolute magnitudes
$M$ using the usual formula:
\begin{equation}
M=m-5(\log_{10}D_L -1)-kcor 
\end{equation}
where $D_L$ is the luminosity distance (in pc) computed with Ned
Wright's cosmology
calculator\footnote{http://www.astro.ucla.edu/wright/CosmoCalc.html}
and $kcor$ is the k-correction. For each cluster, we compute $kcor$
with LePhare as the average value for all the elliptical  galaxy
  templates with a redshift within $\pm 0.05$ of the cluster
redshift.

The error bars on the galaxy counts are computed as follows.
We consider that the errors on the counts along the red sequence (RS) $N_{RS}$
and the field counts $N_{bkg}$ are poissonian. The GLF is defined by:
\begin{equation}
\tilde{N} = \tilde{N}_{RS} - \tilde{N}_{bkg} \mbox{,}
\end{equation}
where  $\tilde{N}_{RS} = \frac{N_{RS}}{A_{RS}}$ is the number of galaxies along 
the RS normalised to 1~deg$^2$ ($A_{RS} = \pi *R_{200}^2$ ) and 
$\tilde{N}_{bkg} = \frac{N_{bkg}}{A_{bkg}}$ is the number of background galaxies normalised to
1~deg$^2$ ($A_{bkg} = 1.38$~deg$^2$ ). 

The error on the galaxy counts normalised to 1~deg$^2$ is then:
\begin{equation}
\tilde{E} = \sqrt{\tilde{E}_{RS}^2 + \tilde{E}_{bkg}^2} \mbox{,}
\end{equation}
\noindent
with $\frac{\tilde{E}_{RS}}{\tilde{N}_{RS}}=\frac{E_{RS}}{N_{RS}}=\frac{1}{\sqrt{N_{RS}}}$ 
and $\frac{\tilde{E}_{bkg}}{\tilde{N}_{bkg}}=\frac{E_{bkg}}{N_{bkg}}=\frac{1}{\sqrt{N_{bkg}}}$
(the relative errors remain the same).

The final error $\tilde{E}$ on the normalised GLFs for individual clusters is therefore:
\footnotesize
\begin{equation}
\tilde{E} = \sqrt{\tilde{E}_{RS}^2 + \tilde{E}_{bkg}^2} = \sqrt{\frac{\tilde{N}_{RS}}{A_{RS}} + \frac{\tilde{N}_{bkg}}{A_{bkg}}}=  \sqrt{\frac{N_{RS}}{A_{RS}^2}+\frac{N_{bkg}}{A_{bkg}^2}}
\end{equation}

\normalsize
We then fit the GLFs with a Schechter function:
\footnotesize
\begin{equation}
\Phi(M)= 0.4\ \ln(10)\Phi^*\times[10^{0.4(M^{*}-M)}]^{\alpha+1}\times\exp(-10^{0.4(M^{*}-M)})
\label{schechter}
\end{equation}

\noindent
\normalsize
where $\Phi^*$ is the normalisation factor, $M^*$ is the absolute magnitude where
the regime changes from bright to faint galaxies and $\alpha$ is the
faint end slope.  The fit is made by minimizing a $\chi ^2$ using
the MINUIT routine.

GLFs are then stacked in mass and redshift bins to improve the quality
of the fits and see if a trend can be found. For this, we follow the
prescription developed by \citet{Colless89}, where the clusters are
normalised to the same solid angle (1~deg$^2$) and to the same richness,
defined as the number of galaxies in a given band up to a certain
limiting magnitude, which we will take to be the 80\% completeness
limit (see Table~\ref{tab:completeness}).  As discussed in e.g. 
\citet{Martinet17}, the Colless stack, although it allows to
maximize the information from the available data as compared to a
fixed number of clusters per bin, presents the caveat that the stack
is dominated by the low-redshift clusters, since these tend to have a
deeper completeness limit. The redshift bins used in this analysis are
however sufficiently thin to study a possible evolution from $z=0.2$
to $z=0.5$.  Following the prescription by \citet{Popesso05b}, the
galaxy number counts in bin $j$ of the stacked GLF are:

\begin{equation}
N_{sj} = \frac{N_{s0}}{Nc_{tot}Nc_j}\sum\limits_{i} \frac{\tilde{N_{ij}}}{N_{i0}} \mbox{,}
\end{equation}
\noindent
where $\tilde{N_{ij}}$ is the number of galaxies in bin $j$ for cluster
  $i$ normalised to 1~deg$^2$, $N_{i0}$ corresponds to the richness
  of cluster $i$, $N_{s0}$ is the sum of the richnesses of all the
  clusters:

\begin{equation}
N_{s0} = \sum\limits_{i} N_{i0} \mbox{,}
\end{equation}

\noindent
$Nc_{tot}$ is the total number of clusters in the stack and $Nc_j$ is
the number of clusters contributing to bin $j$.

The error on $N_{sj}$ is obtained from the Poisson errors
$\tilde{E_{ij}}$ (see above):

\begin{equation}
E_{sj} = \frac{N_{s0}}{m_{tot}m_j}\left [\sum\limits_{i} (\frac{\tilde{E_{ij}}}{N_{i0}})^2 \right ]^{(1/2)} \mbox{.}
\end{equation}

\noindent
The cluster richness is used as a normalisation for the stacks, allowing a
direct comparison of the values of $\Phi ^*$ from one stack to another.

\subsection{Results}

Individual GLFs are first computed for all the clusters, except
  for the three clusters that we eliminate because they only have
  photometric redshifts. Schechter fits are made for both completeness
  limits (90\% and 80\%), but since the results are not very different
  we choose to give the results only for the 80\% completeness limit
  (for which the number of converging fits is slightly larger). The
  parameters of the Schechter fits for the individual clusters are
  given in Appendix~C, Table~C1.  For three clusters, the GLF cannot
  be fit by a Schechter function in any band, so they do not appear in
  Table~C1.  For fifteen clusters, the individual GLF fits are of poor
  quality, with large error bars on the parameters. Out of these, six
  are distant ($z>0.65$), and three have 
   one or several bright
  foreground galaxies close to the cluster center. The remaining six
  clusters are neither particularly distant nor massive, so the reason
  for the poor quality of the GLF fit is unclear.  For most clusters
the GLFs in the u band are too faint to allow a fit by a Schechter
function, so this band will not be discussed.

We want to stress the fact that the minimization procedure used here
to fit the GLFs with Schechter functions gives the $\Phi^*$, $M^*$ and
$\alpha$ parameters with the error bars that we give in the various
tables, but, as we have noted in many of our previous papers, these
error bars are always underestimated. This must be kept in mind when
comparing GLFs and trying to derive conclusions.

We discuss below the GLF Schechter parameters and show the
corresponding figures for stacked clusters (in mass and redshift
bins). 

\subsubsection{GLFs in mass bins}

\begin{table*}
\small
\caption{GLF parameter fits of stacked galaxy luminosity functions for low mass 
($M<7\  10^{13}$~M$_\odot$), medium mass ($7\ 10^{13} \leq M \leq
  10^{14}$~M$_\odot$), and high mass ($M> 10^{14}$~M$_\odot$)
  clusters. Cols.~2--4 correspond to stacking all the clusters with converging fits (stack $a$)
  and Cols.~5--7 to stacking only the clusters for which the errors on the GLF fits are not too large (stack $b$).
  The numbers of clusters included in each stack are indicated in parentheses. In the two cases indicated with
an asterisk, the fits do not converge since the values $M^*_z$ are at their limit value of $-26.0$.}
\begin{tabular}{|c|ccc|ccc|}
\hline
\hline
& \multicolumn{3}{c}{stack $a$} & \multicolumn{3}{c}{stack $b$} \\
            & Low mass (16)   & Medium mass (16) & High mass (12)  & Low mass (12)    & Medium mass (14) & High mass (9)  \\
\hline
$\Phi_g$    &                 & $260\pm 25$      & $152\pm 19$     &                  & $223\pm 22$      & $346\pm 46$    \\
M$^*_g$     &                 & $-23.6\pm 0.1$   & $-25.6\pm 0.2$  &                  & $-23.7\pm 0.1$   & $-23.6\pm 0.2$ \\
$\alpha _g$ &                 & $-1.31\pm 0.02$  & $-1.36\pm 0.02$ &                  & $-1.33\pm 0.02$  & $-1.36\pm 0.02$ \\
\hline
$\Phi_r$    & $178\pm 106$    & $304\pm 15$      & $413\pm 38$     & $150\pm 66$      & $240\pm 17$      & $663\pm 91$     \\
M$^*_r$     & $-25.0\pm 1.5$  & $-24.2\pm 0.1$   & $-24.5\pm 0.1$  & $-24.7\pm 0.7$   & $-24.3\pm 0.1$   & $-23.1\pm 0.1$  \\
$\alpha _r$ & $-1.24\pm 0.05$ & $-1.29\pm 0.01$  & $-1.21\pm 0.02$ & $-1.38\pm 0.04$  & $-1.32\pm 0.01$  & $-1.22\pm 0.04$ \\
\hline
$\Phi_i$    & $413\pm 94$     & $363\pm 16$       & $647\pm 36$     & $232\pm 68$     & $258\pm 15$      & $756\pm 66$     \\
M$^*_i$     & $-23.7\pm 0.2$  & $-24.0\pm 0.1$    & $-24.0\pm 0.1$  & $-24.4\pm 0.4$  & $-24.4\pm 0.1$   & $-23.4\pm 0.1$  \\
$\alpha _i$ & $-1.10\pm 0.06$ & $-1.24\pm 0.01$   & $-1.16\pm 0.02$ & $-1.30\pm 0.04$ & $-1.30\pm 0.01$  & $-1.18\pm 0.02$ \\
\hline   
$\Phi_z$    & $377\pm 110$    & $152\pm 19$       & $968\pm 57$     & $124\pm 11 ^*$     & $112\pm 5 ^*$       & $1085\pm 106$   \\
M$^*_z$     & $-24.0\pm 0.3$  & $-25.6\pm 0.2$    & $-23.9\pm 0.1$  & $-26.0\pm 1.0 ^*$  & $-26.0\pm 0.2 ^*$   & $-23.3\pm 0.1$  \\
$\alpha _z$ & $-1.12\pm 0.08$ & $-1.36\pm 0.02$   & $-1.06\pm 0.02$ & $-1.36\pm 0.03 ^*$ & $-1.38\pm 0.01 ^*$  & $-1.03\pm 0.04$ \\
\hline
\end{tabular}
\label{tab:fitmassbins}
\end{table*}

\begin{figure}   
\includegraphics[width=8cm]{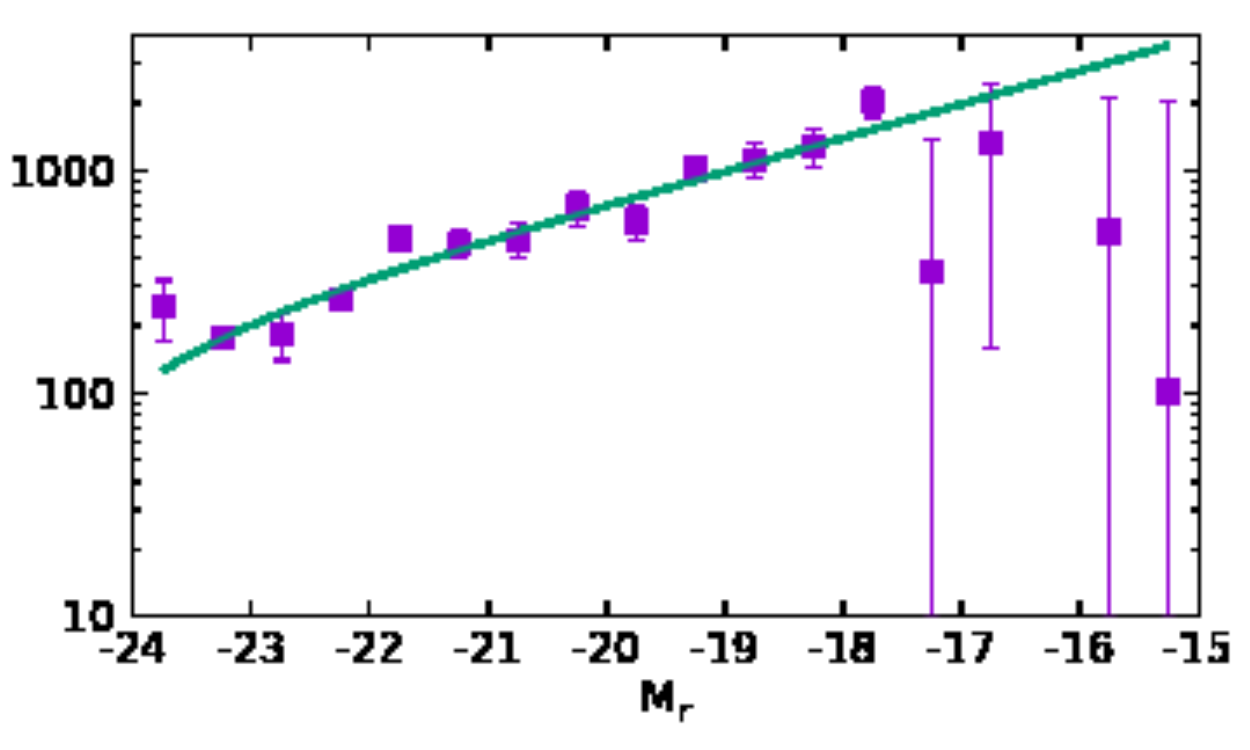}
\includegraphics[width=8cm]{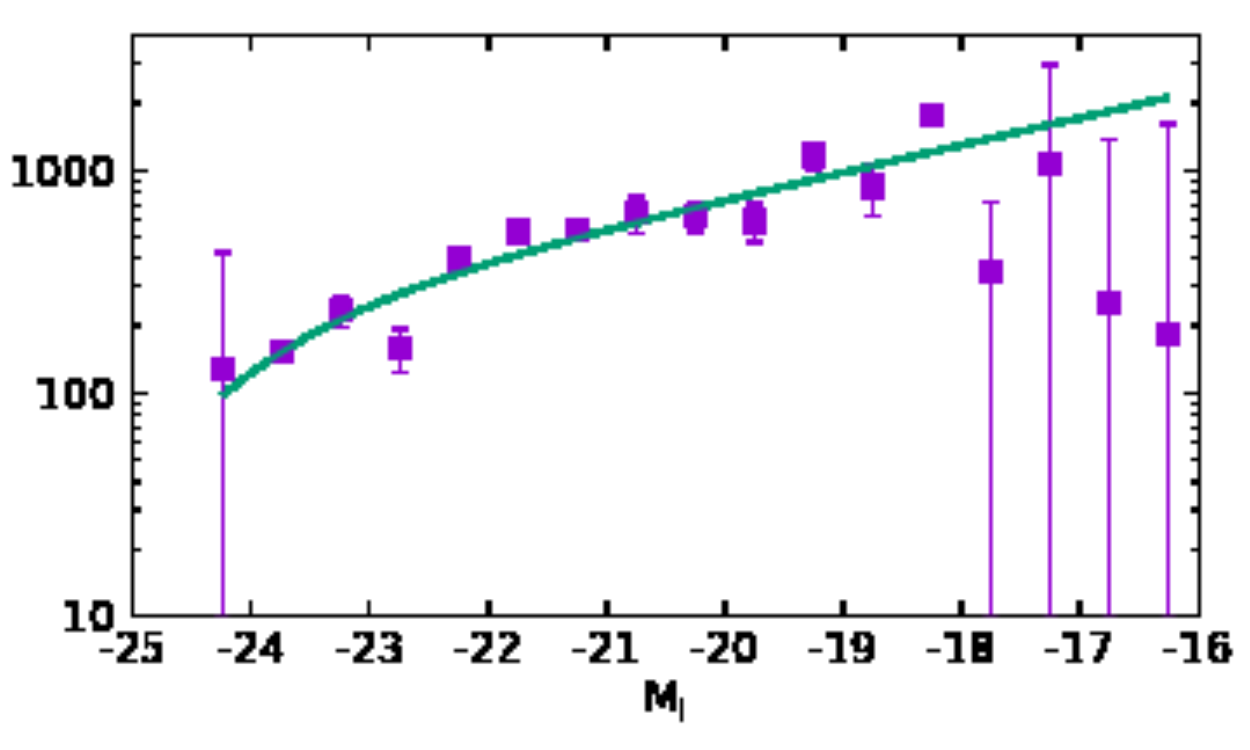}
\includegraphics[width=8cm]{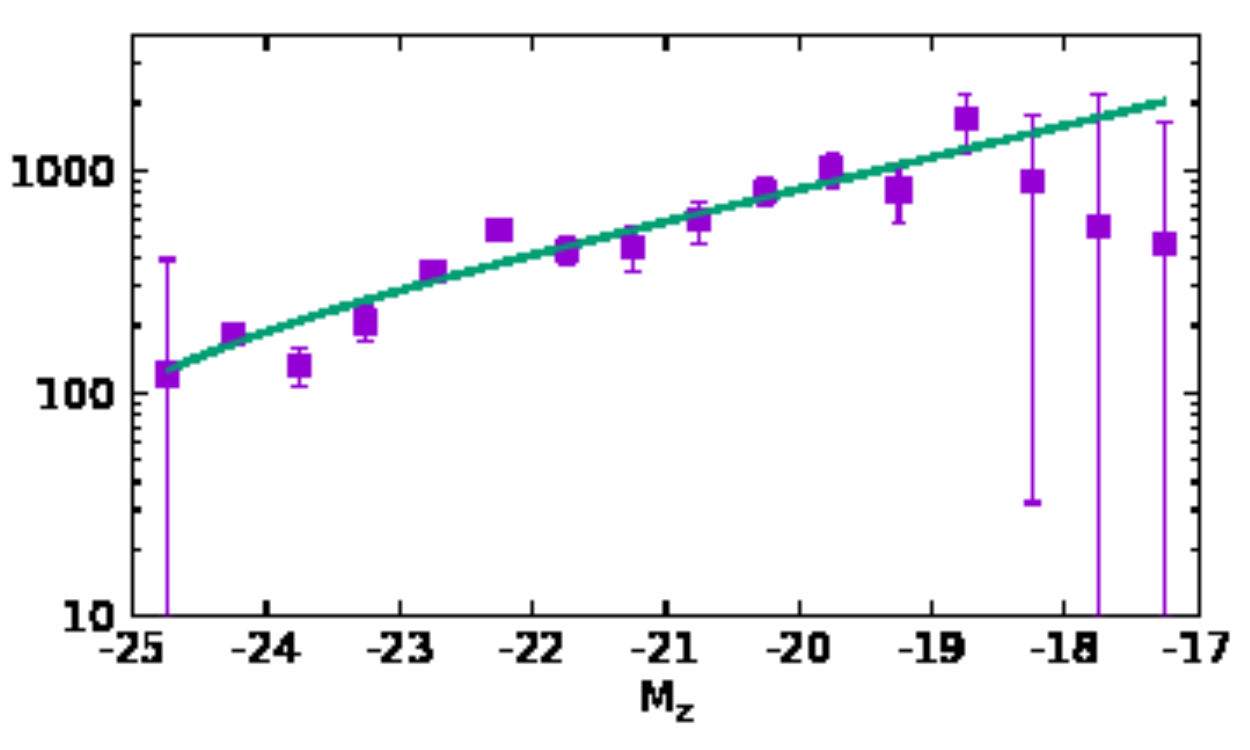}
\caption{GLFs and Schechter fits for the stack of 12 low mass 
($M<7\ 10^{13}$~M$_\odot$) clusters. }
\label{fig:GLF_lowM12}
\end{figure}

\begin{figure}   
\includegraphics[width=8cm]{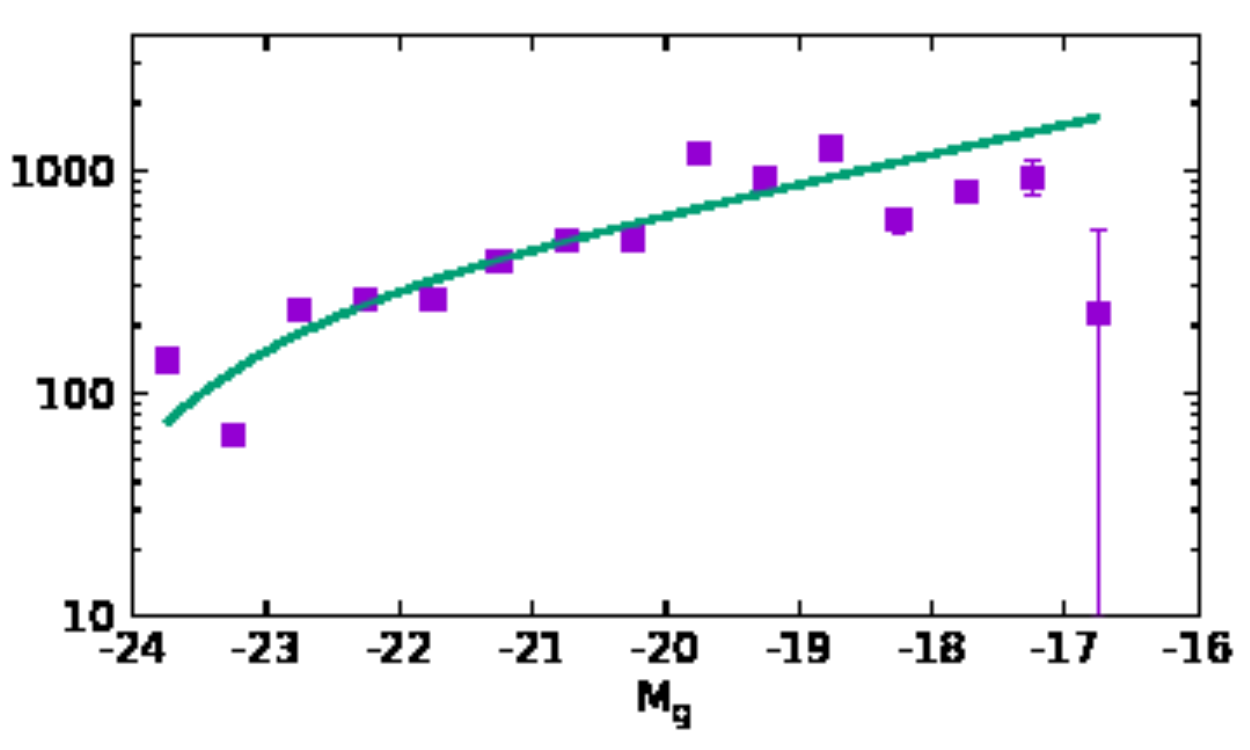}
\includegraphics[width=8cm]{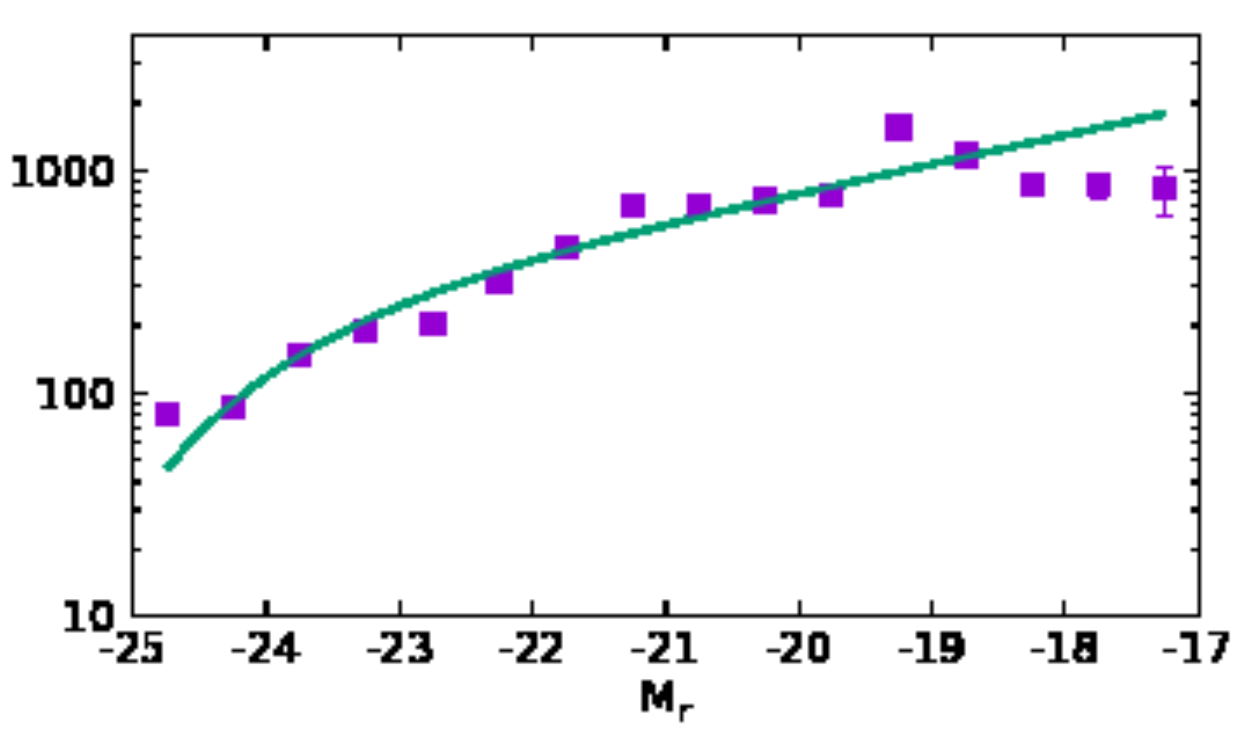}
\includegraphics[width=8cm]{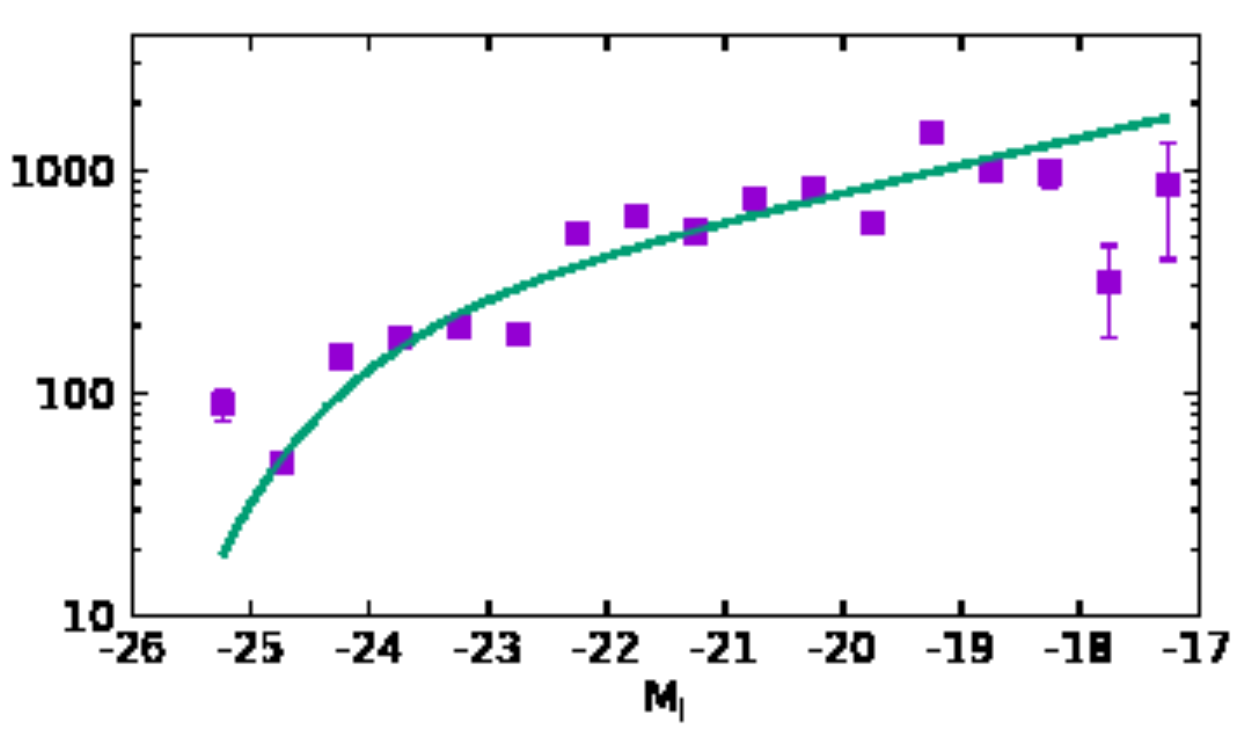}
\includegraphics[width=8cm]{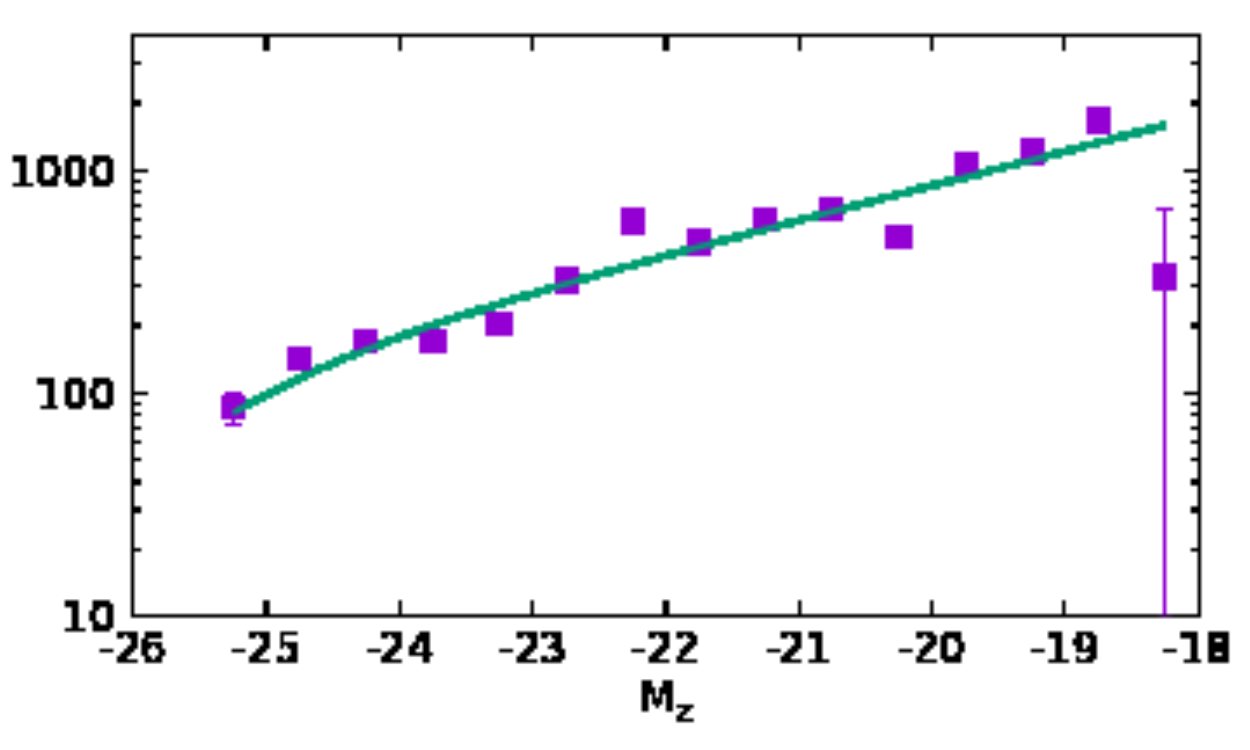}
\caption{GLFs and Schechter fits for the stack of 14 medium 
mass ($7\ 10^{13} \leq M \leq 10^{14}$~M$_\odot$)  clusters. }
\label{fig:GLF_midM14}
\end{figure}

\begin{figure}   
\includegraphics[width=8cm]{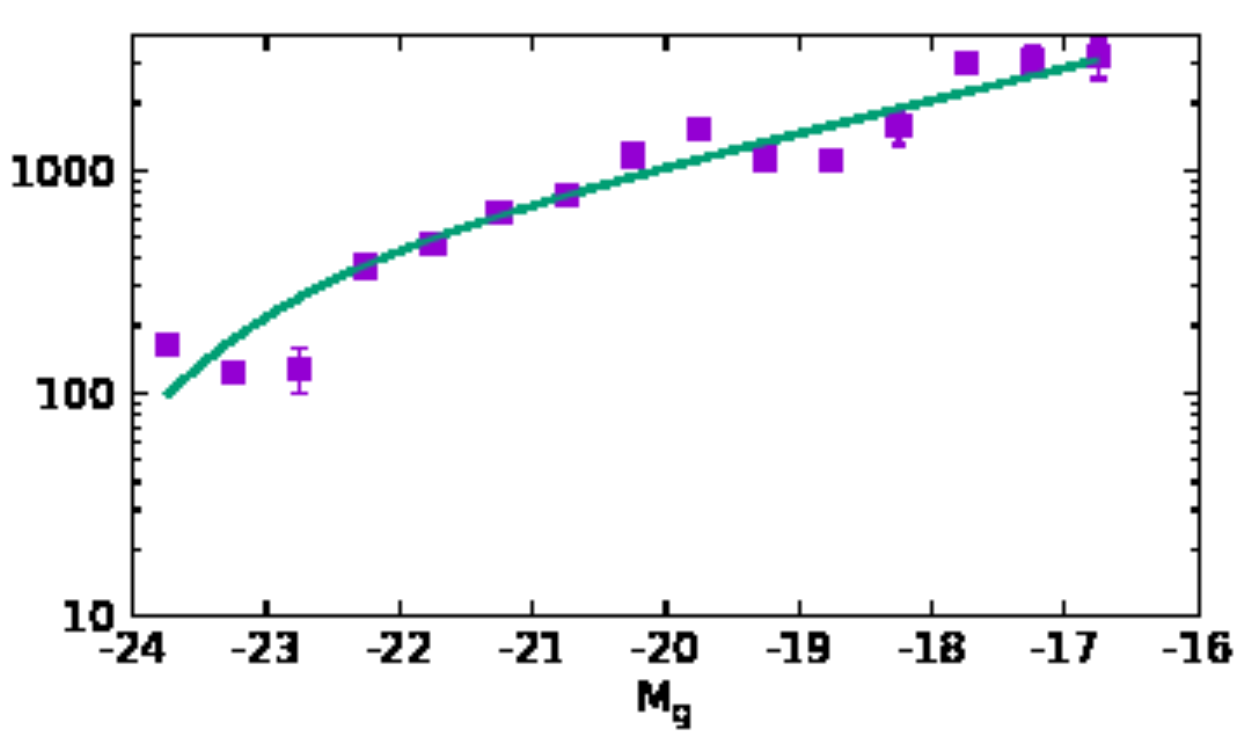}
\includegraphics[width=8cm]{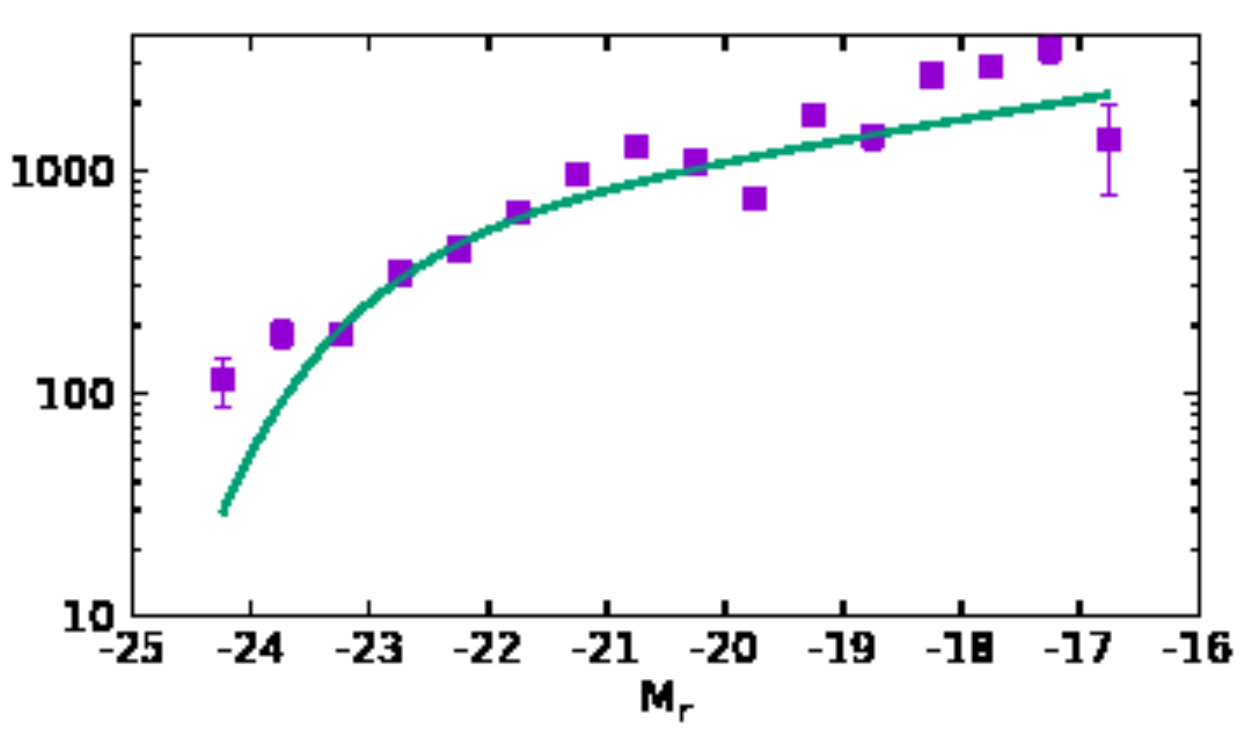}
\includegraphics[width=8cm]{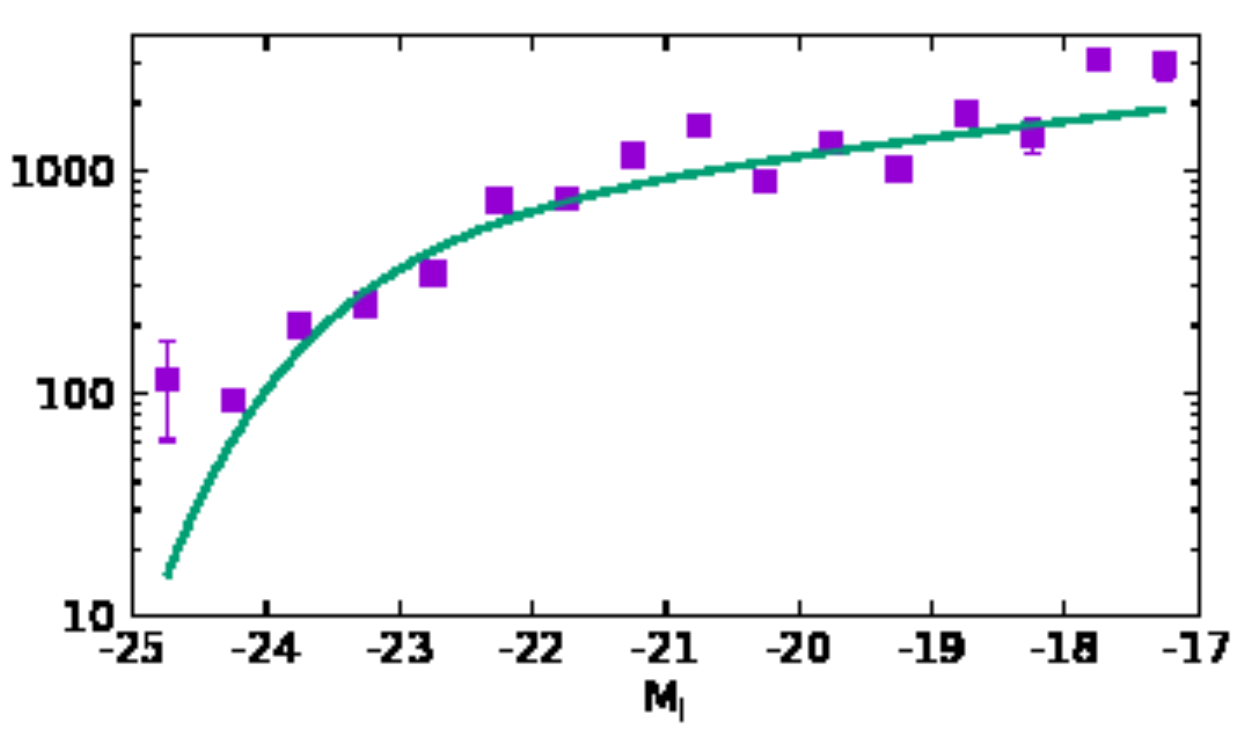}
\includegraphics[width=8cm]{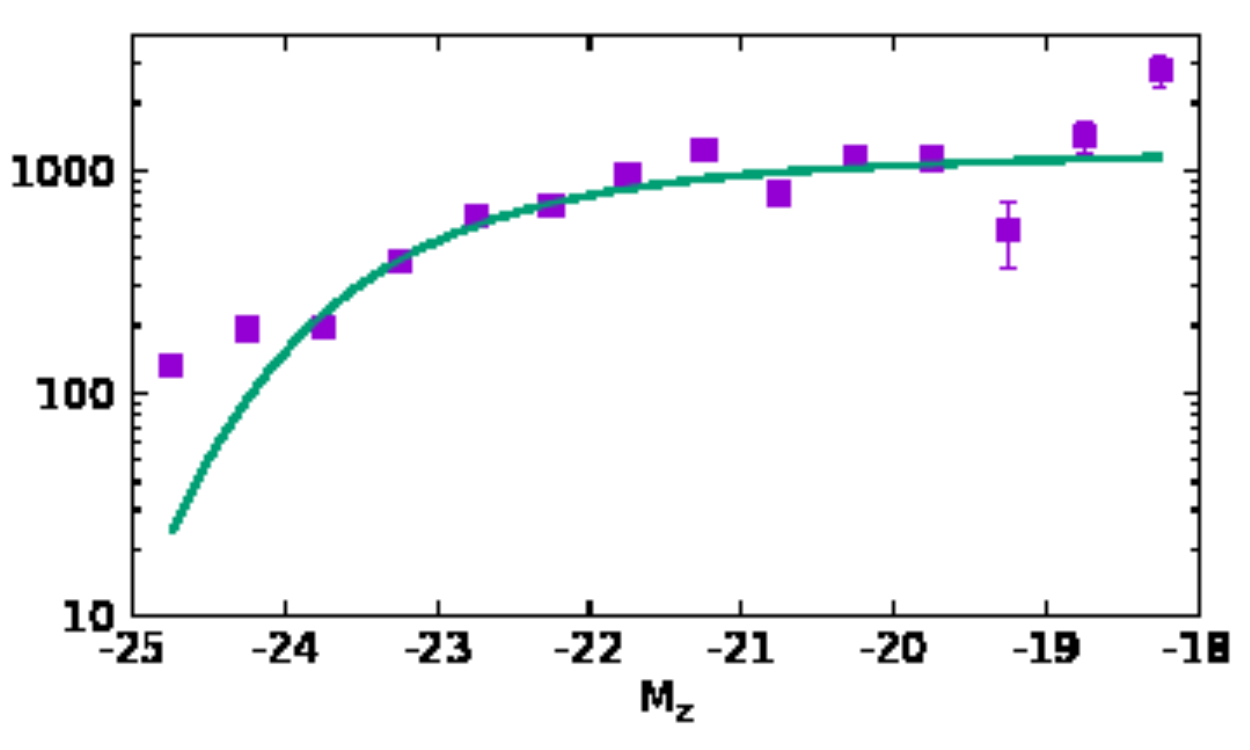}
\caption{GLFs and Schechter fits for the stack of 9 high 
mass  ($M> 10^{14}$~M$_\odot$) clusters. }
\label{fig:GLF_highM9}

\end{figure}

The histogram of cluster masses within $R_{500}$ has been shown in the
previous Section.  To see if we can detect a variation of the
Schechter parameters with cluster mass, we divide our sample into
three mass bins: low mass ($M<7\ 10^{13}$~M$_\odot$), medium mass
($7\ 10^{13} \leq M \leq 10^{14}$~M$_\odot$), and high mass
($M> 10^{14}$~M$_\odot$) clusters. We first include in stack $a$ all
the clusters with converging Schechter fits (44 clusters). These are
distributed as follows: 16 low mass, 16 medium mass and 12 high mass
clusters.  We then try including in stack $b$ only the 35 clusters
with Schechter fits that do not show too large error bars. This gives
12 low mass, 14 medium mass and 9 high mass clusters. 
Clusters belonging to these two stacks are respectively noted with
superscripts $a$ and $b$ in Table~\ref{tab:Sch_param_all}.

The Schechter fit parameters for the stacks in mass bins are given in
Table~\ref{tab:fitmassbins} and the corresponding GLFs are shown
  in Figs.~\ref{fig:GLF_lowM12}, \ref{fig:GLF_midM14} and
  \ref{fig:GLF_highM9}.  We can note that the Schechter fit
parameters are very similar (within error bars) for the two different
stacks for medium and high mass clusters. They differ a little more
for low mass clusters, but these differences are not statistically
significant.

This implies that the GLF fits remain comparable even when a few
clusters with low signal to noise are included, a rather comforting
result.  We do not give the Schechter parameters in the g band for low
mass clusters because they do not converge. For low and medium mass
clusters of stack $b$, we give the Schechter parameters in the z band
as an indication, because the GLFs obtained are ``reasonable'' (see
the bottom of Figs.~\ref{fig:GLF_lowM12} and \ref{fig:GLF_midM14}) but
the $M^*$ parameters are at their limits of $-26.0$, so the fits are
not reliable.  All the other fits converge, but we must keep in
mind the fact that the error bars on the Schechter parameters are
probably underestimated by a factor between 1 and 1.5, based on our
previous experience. Variance from one cluster to another induces
variations in the Schechter fit parameters, and since there are
between 9 and 16 clusters in our stacks the uncertainties on the
parameters are larger than if we were stacking hundreds of clusters 
\citep[see for example the error bars in the Tables of Appendix C in][]{Sarron18}.

For high mass clusters, which are the most reliable, we can see that
the faint end slope $\alpha$ clearly varies with the photometric band:
it smoothly flattens from g to z. This has already been noted for
other clusters such as Coma \citep[see e.g.][
Fig.~13]{Adami07}. Besides this trend, it is difficult to claim any
other significant variation of Schechter fit parameters (given in
Table~\ref{tab:fitmassbins}) with cluster mass.

The absence of a variation of the GLF with cluster mass was already
noted in the extensive study of cluster GLFs made by \citet{Sarron18},
based on a catalogue of 1371 cluster candidates in the Canada France
Hawaii Telescope Legacy Survey (see their Fig.~B.2).  These authors
found that it is only when blue and red galaxies are separated that
the GLFs of blue and red galaxies start showing differences with
cluster mass. Here, we are only studying red galaxies, but our
cluster sample is much smaller than that of \citet{Sarron18}, so we
cannot reach a definite conclusion on the variation of Schechter fit
parameters with cluster mass. 

\subsubsection{GLFs in redshift bins}

The histogram of cluster redshifts is shown in Fig.~\ref{fig:z_hist}.
There are 12 clusters with redshifts $z>0.5$, but for two of them
the galaxy counts are barely above the background so we
could not fit the GLFs and we decided not to include them in the
stack.  We therefore stack the ten remaining clusters (noted with a
$^c$ in Table~\ref{tab:Sch_param_all}) to compute the ``high
redshift'' GLF. As a comparison, we stack ten low redshift clusters
with redshifts $0.2<z<0.3$ to constitute the ``low redshift'' GLF.
These ten clusters were chosen to have a mass distribution
as close as possible as the high redshift sample, to avoid
introducing a possible influence of the cluster mass on the comparison between
high and low redshift clusters.

The results are given in Table~\ref{tab:fitzbins} for the r, i and z
bands (at high redshift, the fit to the stacked GLF in the g band does
not converge).  To avoid including too many figures, we are not
showing the corresponding GLFs, since they are quite similar to those
given in the previous subsection. 

We can see that the faint end slope is flatter at high redshift in the
r and z bands, but not in the i band, so it is difficult to reach any
definite conclusion. A small trend of a flattening faint end slope was
also found by \citet{Sarron18}, as seen in their Fig.~12, but here
also the trend is stronger for blue galaxies, and since 
  we are only studying the GLFs of red galaxies here, we cannot draw
  firm conclusions on the variation of GLF parameters with redshift.

\begin{table}
  \caption{GLF best fit parameters of stacked galaxy luminosity functions for 10 low redshift
    ($0.2<z<0.3$) and 10 high redshift ($z>0.5$) clusters.}
\begin{tabular}{|c|cc|}
\hline
\hline
            & Low redshift (10) & High redshift (10) \\ 
\hline
$\Phi_r$    & $333\pm 43$       & $1717\pm 75$        \\
M$^*_r$    & $-23.1\pm 0.1$    & $-23.3\pm 0.04$   \\
$\alpha _r$ & $-1.26\pm 0.03$   & $-0.92\pm 0.03$  \\
\hline
$\Phi_i$    & $392\pm 45$       & $715\pm 48$        \\
M$^*_i$    & $-23.3\pm 0.2$    & $-24.8\pm 0.1$   \\
$\alpha _i$ & $-1.18\pm 0.03$   & $-1.24\pm 0.01$  \\
\hline   
$\Phi_z$    & $263\pm 37$       & $1481\pm 67$        \\
M$^*_z$    & $-24.2\pm 0.2$    & $-23.68\pm 0.05$   \\
$\alpha _z$ & $-1.19\pm 0.04$   & $-0.98\pm 0.03$  \\
\hline
\end{tabular}
\label{tab:fitzbins}
\end{table}



\section{Conclusions}

We have presented a detailed analysis of the cluster sample published
from our cluster survey, the 3XMM/SDSS Stripe 82 galaxy cluster
survey. Our study includes 54 clusters in a redshift range from 0.05
to 1.2 (51 spectroscopic redshifts and 3 photometric redshifts).

We first determined the X-ray temperatures and luminosities of 45
clusters through spectral fitting of their spectra, in an aperture of
300 kpc. The X-ray data quality of the remaining systems did not allow
a spectral analysis. The X-ray temperatures are in the range
[1.0-8.0]~keV and the X-ray luminosities in an aperture of radius
300~kpc are in the range [1.0-104]$10^{42}$~erg~s$^{-1}$. For 37
clusters with good quality X-ray data, we investigated the
${\rm L_X-T_X}$ relation, and found a best fit slope of
$3.01\pm 0.51$, similar to values published in the literature. We also found a good agreement between the intercept of our relation with those values derived for other relations based on different cluster samples.
This shows that our sample is representative of typical cluster samples,
with no obvious bias.

We then investigated some optical properties of the cluster galaxies.
First, we computed the fraction of early and late type galaxies as a
function of cluster mass and distance to the cluster X-ray centre.  We
observe no strong variation of the fraction of early and late type
galaxies with cluster mass, except for the most massive clusters,
which contain a somewhat larger fraction of late-type galaxies. This
may be explained by the fact that more massive clusters accrete more
late-type galaxies in their outskirts. As expected, we found a very
large (close to 80\%) fraction of early type galaxies in the innermost
radial bin, decreasing to 50\% at radii above 1.3 Mpc, while the fraction of
late type galaxies increases with radius and becomes larger than 50\%
around 1.3 Mpc. We found no significant variations of the fractions of 
early and late type galaxies as a function of the number of galaxies within $R_{200}$.

Second, we investigated the galaxy luminosity functions (GLFs) in the
five ugriz bands for red sequence selected galaxies.  We limited our study of the individual GLFs to 36
clusters. For the few clusters with GLF fits in the $u$ band, the
faint end slope tends to be steeper than in the other bands, a trend
already noted in other studies \citep[e.g.][and references
therein]{Boue08}. However, in view of the large error bars on the
Schechter parameters in the $u$ band, it is difficult to confirm this
trend. For a given cluster, the faint end slopes in the other bands
are all similar within the error bars.

We then stacked the GLFs in the griz bands in three mass bins and two
redshift bins.  The Schechter fit parameters are very similar for the
stacks for medium and high mass clusters. They are slightly different
for low mass clusters, but this may just be due to the lower quality
of the GLF fits for low mass clusters.  For high mass clusters (the
most reliable), the faint end slope varies with the photometric band,
smoothly flattening from g to z, as previously noted for other
clusters such as Coma \citet{Adami07}. Besides this trend, and keeping
in mind the fact that the error bars on the Schechter parameters are
most probably underestimated, it is difficult to claim any significant
variation of Schechter fit parameters with cluster mass, in agreement
with \citet{Sarron18}.

The comparison of the GLFs stacked in two redshift bins with
comparable mass distributions shows that the faint end slope is
flatter at high redshift ($z>0.5$) in the r and z bands, but not in
the i band, so it is difficult to reach any definite conclusion on the
variation with redshift, also in agreement with \citet{Sarron18}.

Twenty clusters of the present sample are studied for the first time
in X-rays, and the 54 clusters of this sample are studied for the
first time in the optical range. Altogether, our cluster sample
appears to have X-ray and optical properties which are representative
of ``average'' cluster properties, and can therefore be added to other
cluster samples to increase, for example, the statistics on X-ray or
optical cluster properties.



\section*{Acknowledgements}
This work was supported by the Egyptian Science and Technology Development Fund (STDF) and the French Institute in Egypt (IFE) in cooperation with the Institut d'Astrophysique de Paris (IAP), France. F. D. acknowledges long-term support from CNES. I.M. acknowledges support from the Spanish Ministry of Economy and Competitiveness through grants AYA2013-42227-P and AYA2016-76682-C3-1-P.
We are very grateful to Nicolas Martinet and Florian Sarron for discussions on GLFs. We also thank the referee, Rupal Mittal, and the anonymous one for 
their numerous constructive comments and suggestions.

This research has made use of data obtained from the 3XMM XMM-Newton serendipitous source catalogue compiled by the 10 institutes of the XMM-Newton Survey Science Centre selected by ESA. This work is based on observations obtained with XMM-Newton, an ESA science mission with instruments and contributions directly funded by ESA Member States and the USA (NASA). 
This research has made use of the NASA/IPAC Extragalactic Database (NED) which is operated by the Jet Propulsion Laboratory, California Institute of Technology, under contract with the National Aeronautics and Space Administration (NASA).

Funding for SDSS-III has been provided by the Alfred P. Sloan 
Foundation, the Participating Institutions, the National Science Foundation,
and the U.S. Department of Energy. The SDSS-III web site is 
\url{http://www.sdss3.org/}. 
SDSS-III is managed by the Astrophysical Research 
Consortium for the Participating Institutions of the SDSS-III Collaboration 
including the University of Arizona, the Brazilian Participation Group, 
Brookhaven National Laboratory, University of Cambridge, University of 
Florida, the French Participation Group, the German Participation Group, 
the Instituto de Astrofisica de Canarias, the Michigan State/Notre 
Dame/JINA Participation Group, Johns Hopkins University, Lawrence Berkeley 
National Laboratory, Max Planck Institute for Astrophysics, New Mexico State 
University, New York University, Ohio State University, Pennsylvania State 
University, University of Portsmouth, Princeton University, the Spanish 
Participation Group, University of Tokyo, University of Utah, Vanderbilt 
University, University of Virginia, University of Washington, and Yale 
University.





\bibliographystyle{mnras}
\bibliography{refbib_papers} 



\appendix

\section{X-ray parameters for the cluster sample used in the 
${\rm L_X-T_X}$ relation}

The X-ray parameters of our galaxy cluster sample are measured from
spectral fits to cluster spectra extracted from XMM-Newton EPIC (MOS1,
MOS2, pn) observations. Table~\ref{tab:xray-cat} lists the X-ray
parameters of the 37 systems used to investigate the ${\rm L_X-T_X}$
relation. The table columns are: DETID: Detection number in the
3XMM-DR5 catalogue, 3XMM Name: IAU name of the X-ray source, RA: right
ascension of X-ray detection in degrees (J2000), Dec: declination of
detection in degrees (J2000), OBSID: XMM observation identification
number, z: galaxy cluster redshift (note: all the clusters of the sample have spectroscopic redshifts except for 3XMM J213340.8-003841 that has only a photometric redshift), $R_{300~kpc}$: spectrum extraction
radius (300 kpc) in arcsec, $R_{500}$ radius in arcsec computed in the present work (see Section 3.2), 
nH: Galactic hydrogen column density, kT: X-ray
temperature in [0.5-2.0] keV within $R_{300~kpc}$, nekT and pekT:
negative and positive errors on kT, ekT: average error on kT, Lx:
aperture ($R_{300~kpc}$) X-ray luminosity in [0.5-2.0] keV
($10^{42}$~erg~s$^{-1}$), neLx and peLx: negative and positive errors
in Lx, eLx:  average error on Lx, $L_{500}$ and $eL_{500}$: X-ray
bolometric luminosity and its error within $R_{500}$
($10^{42}$~erg~s$^{-1}$).

\begin{landscape}
\begin{table}
  \caption{\label{tab:xray-cat} X-ray parameters of our galaxy cluster sample (37 systems) that are used to plot the ${\rm L_X-T_X}$ relation.}
 {\scriptsize
\begin{tabular}{|r|l|r|r|r|r|r|r|r|r|r|r|r|r|r|r|r|r|r|}
\hline
  \multicolumn{1}{|c|}{DETID} &
  \multicolumn{1}{c|}{3XMM name} &
  \multicolumn{1}{c|}{RA} &
  \multicolumn{1}{c|}{Dec} &
  \multicolumn{1}{c|}{OBSID} &
  \multicolumn{1}{c|}{z} &
  \multicolumn{1}{c|}{$R_{300~kpc}$} &
  \multicolumn{1}{c|}{$R_{500}$} &
  \multicolumn{1}{c|}{nH} &
  \multicolumn{1}{c|}{kT} &
  \multicolumn{1}{c|}{nekT} &
  \multicolumn{1}{c|}{pekT} &
  \multicolumn{1}{c|}{ekT} &
  \multicolumn{1}{c|}{Lx} &
  \multicolumn{1}{c|}{neLx} &
  \multicolumn{1}{c|}{peLx} &
  \multicolumn{1}{c|}{eLx} &
  \multicolumn{1}{c|}{$L_{500}$} &
  \multicolumn{1}{c|}{$eL_{500}$} \\
\hline
\hline
104037603010094 & J001115.5+005152 &   2.81470 &   0.86462 & 0403760301 & 0.3622 &  59.4 &  93.3 & 0.027 & 2.32 & 0.51 & 0.85 & 0.68 &   3.3 & 0.4 & 0.5 & 0.5 &   9.0 &  1.1 \\
102036901010028 & J003838.0+004351 &   9.65851 &   0.73108 & 0203690101 & 0.6955 &  42.1 &  77.0 & 0.020 & 4.50 & 0.99 & 1.45 & 1.22 &  22.2 & 1.7 & 1.9 & 1.8 &  83.2 &  5.6 \\
102036901010085 & J003840.3+004747 &   9.66813 &   0.79659 & 0203690101 & 0.5553 &  46.6 &  82.0 & 0.020 & 3.05 & 0.77 & 1.09 & 0.93 &  12.4 & 1.4 & 1.3 & 1.3 &  38.2 &  3.0 \\
102036901010023 & J003922.4+004809 &   9.84359 &   0.80277 & 0203690101 & 0.4145 &  54.7 & 108.1 & 0.020 & 3.79 & 0.51 & 0.52 & 0.52 &  12.5 & 0.6 & 0.6 & 0.6 &  45.9 &  1.0 \\
102036901010017 & J003942.2+004533 &   9.92597 &   0.75926 & 0203690101 & 0.4156 &  54.6 & 104.4 & 0.019 & 2.36 & 0.32 & 0.35 & 0.34 &  13.5 & 0.9 & 0.7 & 0.8 &  37.5 &  1.3 \\
100900702010087 & J004231.0+005112 &  10.62930 &   0.85336 & 0090070201 & 0.1579 & 110.0 & 182.3 & 0.018 & 1.70 & 0.16 & 0.21 & 0.18 &   2.5 & 0.2 & 0.2 & 0.2 &   6.1 &  0.4 \\
100900702010056 & J004252.5+004300 &  10.71892 &   0.71692 & 0090070201 & 0.2697 &  72.6 & 133.8 & 0.018 & 2.63 & 0.48 & 0.75 & 0.61 &   5.8 & 0.5 & 0.4 & 0.5 &  17.3 &  1.1 \\
100900702010050 & J004334.1+010107 &  10.89187 &   1.01811 & 0090070201 & 0.2000 &  90.9 & 172.6 & 0.018 & 1.60 & 0.10 & 0.09 & 0.09 &   7.0 & 0.4 & 0.4 & 0.4 &  16.3 &  0.7 \\
100900702010052 & J004350.6+004731 &  10.96114 &   0.79216 & 0090070201 & 0.4754 &  50.5 & 100.4 & 0.018 & 3.76 & 0.70 & 1.13 & 0.92 &  17.1 & 1.3 & 1.5 & 1.4 &  59.6 &  3.6 \\
103035622010028 & J004401.4+000644 &  11.00583 &   0.11226 & 0303562201 & 0.2187 &  84.8 & 182.4 & 0.017 & 2.59 & 0.46 & 0.67 & 0.56 &  12.2 & 0.9 & 1.1 & 1.0 &  37.8 &  1.5 \\
103031104010030 & J005546.1+003839 &  13.94249 &   0.64422 & 0303110401 & 0.0665 & 235.3 & 403.3 & 0.028 & 1.30 & 0.05 & 0.05 & 0.05 &   2.6 & 0.2 & 0.1 & 0.1 &   5.6 &  0.2 \\
106053911010001 & J012023.3-000444 &  20.09717 &  -0.07908 & 0605391101 & 0.0780 & 203.3 & 438.2 & 0.034 & 1.64 & 0.03 & 0.03 & 0.03 &   9.6 & 0.1 & 0.2 & 0.2 &  24.1 &  0.5 \\
101016402010005 & J015917.1+003010 &  29.82144 &   0.50300 & 0101640201 & 0.3820 &  57.4 & 161.2 & 0.023 & 2.91 & 0.33 & 0.31 & 0.32 & 104.3 & 3.5 & 3.9 & 3.7 & 360.2 &  8.5 \\
101016402010018 & J020019.2+001932 &  30.08002 &   0.32564 & 0101640201 & 0.6825 &  42.4 &  84.3 & 0.023 & 1.90 & 0.25 & 0.38 & 0.31 &  58.5 & 6.8 & 7.8 & 7.3 & 132.8 & 15.1 \\
106524006010012 & J022825.8+003203 &  37.10780 &   0.53441 & 0652400601 & 0.3952 &  56.3 & 125.5 & 0.023 & 3.40 & 0.25 & 0.39 & 0.32 &  26.2 & 0.8 & 0.9 & 0.9 &  90.4 &  0.3 \\
106524006010017 & J022830.5+003032 &  37.12738 &   0.50907 & 0652400601 & 0.7214 &  41.5 &  85.2 & 0.023 & 6.31 & 1.07 & 1.69 & 1.38 &  38.4 & 2.2 & 2.9 & 2.5 & 188.9 & 11.9 \\
106524007010008 & J023026.7+003733 &  37.61157 &   0.62602 & 0652400701 & 0.8600 &  39.0 &  83.5 & 0.021 & 7.96 & 1.46 & 2.56 & 2.01 &  72.4 & 3.7 & 3.9 & 3.8 & 419.5 &  9.2 \\
106064313010011 & J025846.5+001219 &  44.69388 &   0.20555 & 0606431301 & 0.2589 &  74.8 & 167.7 & 0.065 & 3.73 & 0.95 & 1.53 & 1.24 &  14.2 & 1.4 & 1.4 & 1.4 &  56.8 &  3.7 \\
106064313010004 & J025932.5+001353 &  44.88574 &   0.23161 & 0606431301 & 0.1920 &  93.9 & 223.2 & 0.067 & 3.43 & 0.58 & 0.85 & 0.71 &  16.6 & 1.1 & 1.1 & 1.1 &  64.5 &  1.6 \\
100411701010097 & J030145.7+000323 &  45.44072 &   0.05659 & 0041170101 & 0.6900 &  42.2 &  67.0 & 0.070 & 1.71 & 0.32 & 0.98 & 0.65 &  15.0 & 2.5 & 4.7 & 3.6 &  33.4 &  7.5 \\
100411701010074 & J030212.1-000132 &  45.55054 &  -0.02579 & 0041170101 & 1.1900 &  36.2 &  53.3 & 0.071 & 3.87 & 0.69 & 0.98 & 0.84 &  33.1 & 3.2 & 3.3 & 3.2 & 140.0 & 13.2 \\
100411701010113 & J030212.1+001107 &  45.55082 &   0.18536 & 0041170101 & 0.6523 &  43.2 &  59.6 & 0.069 & 1.83 & 0.25 & 0.57 & 0.41 &   5.4 & 0.8 & 1.5 & 1.1 &  12.0 &  2.6 \\
100411701010112 & J030317.5+001245 &  45.82296 &   0.21272 & 0041170101 & 0.5900 &  45.2 &  66.4 & 0.068 & 1.58 & 0.27 & 0.30 & 0.28 &   7.0 & 1.1 & 1.2 & 1.1 &  14.1 &  2.3 \\
101426101010024 & J030614.1-000540 &  46.55923 &  -0.09474 & 0142610101 & 0.4249 &  53.9 & 103.0 & 0.063 & 2.15 & 0.23 & 0.31 & 0.27 &  14.0 & 0.7 & 1.3 & 1.0 &  38.9 &  1.9 \\
101426101010022 & J030617.3-000836 &  46.57206 &  -0.14361 & 0142610101 & 0.1093 & 150.5 & 268.5 & 0.064 & 1.68 & 0.08 & 0.25 & 0.16 &   3.4 & 0.1 & 0.2 & 0.2 &   8.2 &  0.2 \\
101426101010059 & J030633.1-000350 &  46.63804 &  -0.06408 & 0142610101 & 0.1235 & 135.3 & 193.4 & 0.063 & 1.46 & 0.12 & 0.16 & 0.14 &   1.1 & 0.1 & 0.1 & 0.1 &   2.2 &  0.2 \\
102011201010042 & J030637.3-001801 &  46.65570 &  -0.30054 & 0201120101 & 0.4576 &  51.6 &  90.9 & 0.063 & 2.54 & 0.51 & 0.82 & 0.66 &   8.9 & 1.3 & 1.5 & 1.4 &  26.3 &  3.5 \\
104023202010027 & J033446.2+001710 &  53.69279 &   0.28618 & 0402320201 & 0.3261 &  63.6 & 130.1 & 0.070 & 2.49 & 0.47 & 0.92 & 0.70 &  13.3 & 1.5 & 2.4 & 1.9 &  40.7 &  4.2 \\
101349209010028 & J035416.9-001003 &  58.57060 &  -0.16751 & 0134920901 & 0.2100 &  87.5 & 190.9 & 0.117 & 4.60 & 1.05 & 1.68 & 1.37 &   7.8 & 0.6 & 0.8 & 0.7 &  40.1 &  2.4 \\
103048012010021 & J213340.8-003841 & 323.41996 &  -0.64481 & 0304801201 & 0.2110 &  87.2 & 197.3 & 0.036 & 3.60 & 0.50 & 0.67 & 0.59 &  13.4 & 0.6 & 0.8 & 0.7 &  50.6 &  1.4 \\
106553468400021 & J221211.0-000833 & 333.04618 &  -0.14275 & 0655346840 & 0.3643 &  59.2 & 119.6 & 0.044 & 3.60 & 0.49 & 2.41 & 1.45 &  17.1 & 2.1 & 2.6 & 2.4 &  43.4 &  4.8 \\
106553468390009 & J221422.1+004712 & 333.59226 &   0.78680 & 0655346839 & 0.3202 &  64.4 & 132.2 & 0.035 & 4.28 & 1.28 & 2.17 & 1.72 &   9.9 & 1.6 & 1.5 & 1.6 &  40.7 &  5.2 \\
106700202010013 & J222144.0-005306 & 335.43347 &  -0.88513 & 0670020201 & 0.3353 &  62.5 & 119.8 & 0.047 & 2.10 & 0.38 & 0.48 & 0.43 &  10.6 & 1.0 & 1.7 & 1.4 &  28.2 &  2.4 \\
106524010010043 & J232809.0+001116 & 352.03771 &   0.18778 & 0652401001 & 0.2780 &  71.1 & 125.9 & 0.041 & 3.25 & 0.72 & 1.40 & 1.06 &   4.1 & 0.4 & 0.3 & 0.3 &  14.0 &  1.0 \\
106524011010056 & J232925.6+000554 & 352.35668 &   0.09849 & 0652401101 & 0.4021 &  55.7 &  86.4 & 0.041 & 2.35 & 0.51 & 0.93 & 0.72 &   3.6 & 0.4 & 0.5 & 0.4 &   9.7 &  1.1 \\
106524014010034 & J233138.1+000738 & 352.90912 &   0.12725 & 0652401401 & 0.2238 &  83.4 & 138.0 & 0.041 & 1.30 & 0.04 & 0.04 & 0.04 &   3.9 & 0.1 & 0.2 & 0.2 &   7.5 &  0.2 \\
106524013010039 & J233328.1-000123 & 353.36739 &  -0.02308 & 0652401301 & 0.5120 &  48.5 &  94.3 & 0.039 & 6.14 & 1.85 & 2.92 & 2.39 &  11.8 & 1.4 & 1.4 & 1.4 &  59.7 &  6.1 \\
\hline
\end{tabular}
}
\end{table}
\end{landscape}



\section{Brightest cluster galaxies (BCGs)}

We give in Table~\ref{tab:BCGs} the list of the BCGs for 53 of our 54
clusters with their positions, spectroscopic and photometric
redshifts, and magnitudes in the five bands (we could not identify the BCG of cluster 3XMM J030637.3-001801). 

\begin{table*}
\caption{List of BCGs for 53 clusters. 
The columns are: 1)~cluster name, 2) and 3)~RA and Dec (J2000.0) of
the BCG, 4)~spectroscopic redshift (zspec), 5)~photometric redshift (zphot), 6)~ to 10) observed magnitudes in the u, g, r, i, and z bands. Quantities that could not be measured are noted as 99.99. } 
\begin{tabular}{crrrrrrrrr}
\hline
\hline
Cluster (3XMM)   & RA        & Dec       & zspec    & zphot    & u      & g      & r      & i      & z \\
\hline
J001115.5+005152 &   2.81328 &   0.86553 &  0.36470 &  0.39719 & 99.990 & 20.416 & 18.327 & 17.911 & 16.970 \\
J001737.3-005240 &   4.40162 &  -0.91354 &  0.21677 &  0.19107 & 20.755 & 19.043 & 17.721 & 17.014 & 16.986 \\
J002223.3+001201 &   5.59128 &   0.19003 &  0.27911 &  0.26268 & 99.990 & 19.392 & 17.838 & 17.293 & 16.971 \\
J002314.4+001200 &   5.81153 &   0.19945 &  0.25966 &  0.25256 & 22.045 & 19.011 & 17.492 & 17.064 & 16.810 \\
J002928.6-001250 &   7.36553 &  -0.19710 &  0.06859 &  0.08381 & 19.606 & 18.344 & 18.018 & 17.809 & 17.751 \\
J003838.0+004351 &   9.65987 &   0.73585 &  0.69818 &  0.90131 & 99.990 & 99.990 & 20.210 & 19.062 & 18.580 \\
J003840.3+004747 &   9.67856 &   0.81097 & 99.99000 &  0.54099 & 99.990 & 22.280 & 21.291 & 20.941 & 99.990 \\
J003922.4+004809 &   9.84172 &   0.80543 &  0.41833 &  0.38856 & 99.990 & 21.115 & 19.169 & 18.456 & 18.197 \\
J003942.2+004533 &   9.93734 &   0.73325 &  0.41513 &  0.37082 & 99.990 & 21.538 & 19.790 & 19.062 & 18.781 \\
J004231.0+005112 &  10.60627 &   0.86424 &  0.16679 &  0.11600 & 19.680 & 18.186 & 17.479 & 17.124 & 16.918 \\
J004252.5+004300 &  10.60627 &   0.86424 &  0.16679 &  0.11600 & 99.990 & 99.990 & 17.479 & 99.990 & 99.990 \\
J004334.1+010107 &  10.83817 &   0.99669 &  0.19777 &  0.18827 & 99.990 & 18.418 & 17.064 & 16.613 & 16.311 \\
J004350.6+004731 &  10.95829 &   0.78839 &  0.47578 &  0.48996 & 99.990 & 21.569 & 19.451 & 18.496 & 18.250 \\
J004401.4+000644 &  11.00534 &   0.11337 &  0.21971 &  0.22075 & 21.353 & 18.718 & 17.058 & 16.732 & 16.571 \\
J005546.1+003839 &  13.93826 &   0.65065 &  0.06991 &  0.08277 & 19.733 & 18.247 & 17.202 & 17.043 & 16.808 \\
J005608.9+004106 &  14.01921 &   0.64810 &  0.06961 &  0.08748 & 20.245 & 18.433 & 17.542 & 17.192 & 17.000 \\
J010606.7+004925 &  16.55266 &   0.87025 &  0.26546 &  0.27514 & 99.990 & 19.558 & 18.084 & 17.686 & 17.368 \\
J010610.0+005108 &  20.08572 &  -0.08110 &  0.08188 &  0.10104 & 20.725 & 18.602 & 17.637 & 17.288 & 17.002 \\
J012023.3-000444 &  29.82316 &   0.51869 &  0.38400 &  0.37363 & 22.979 & 20.848 & 19.051 & 18.524 & 18.114 \\
J015917.1+003010 &  29.95954 &   0.27864 & 99.99000 &  0.77416 & 99.990 & 22.424 & 22.142 & 21.407 & 99.990 \\
J015953.1+001659 &  30.08100 &   0.32492 &  0.68247 &  0.43936 & 99.990 & 22.433 & 20.406 & 18.922 & 18.787 \\
J020019.2+001932 &  32.55104 &  -0.24733 &  0.28275 &  0.25294 & 99.990 & 19.108 & 17.678 & 17.143 & 16.835 \\
J021012.6-001439 &  32.72601 &  -0.39254 &  0.31787 &  0.32762 & 99.990 & 20.050 & 18.347 & 17.764 & 17.485 \\
J021045.8-002156 &  37.12971 &   0.52598 &  0.40177 &  0.66653 & 21.015 & 20.748 & 20.101 & 19.653 & 19.256 \\
J022825.8+003203 &  37.12971 &   0.52598 &  0.40177 &  0.66653 & 21.015 & 20.748 & 20.101 & 19.653 & 19.256 \\
J022830.5+003032 &  37.61845 &   0.59797 & 99.99000 &  0.87046 & 99.990 & 24.132 & 23.021 & 99.990 & 99.990 \\
J023026.7+003733 &  37.74370 &   0.73723 &  0.47402 &  0.48463 & 99.990 & 22.888 & 21.162 & 20.461 & 19.997 \\
J023058.5+004327 &  44.66437 &   0.16252 &  0.26216 &  0.23974 & 99.990 & 19.764 & 18.490 & 17.952 & 17.629 \\
J025846.5+001219 &  44.87015 &   0.26878 &  0.19441 &  0.27494 & 23.168 & 20.798 & 19.517 & 19.061 & 18.688 \\
J025932.5+001353 &  45.42060 &   0.05867 & 99.99000 &  0.73122 & 99.990 & 99.990 & 22.804 & 21.525 & 20.901 \\
J030145.7+000323 &  45.52695 &  -0.00764 & 99.99000 &  0.68029 & 99.990 & 99.990 & 22.419 & 21.245 & 99.990 \\
J030205.6-000001 &  45.57111 &  -0.03033 & 99.99000 &  0.69769 & 99.990 & 99.990 & 21.700 & 19.924 & 19.354 \\
J030212.1-000132 &  45.54819 &   0.18750 &  0.65228 &  0.64433 & 99.990 & 99.990 & 20.707 & 19.431 & 19.280 \\
J030212.1+001107 &  45.82090 &   0.20803 &  0.60487 &  0.58124 & 99.990 & 22.174 & 20.454 & 19.204 & 18.898 \\
J030317.5+001245 &  46.55886 &  -0.09439 &  0.42488 &  0.38512 & 99.990 & 20.605 & 18.797 & 18.033 & 17.666 \\
J030614.1-000540 &  46.63088 &  -0.23376 &  0.11949 &  0.12933 & 19.872 & 17.787 & 16.824 & 16.384 & 16.064 \\
J030617.3-000836 &  46.66190 &  -0.04452 &  0.11252 &  0.17952 & 99.990 & 20.712 & 19.831 & 19.327 & 19.004 \\
J030633.1-000350 &  46.66190 &  -0.04452 &  0.11252 &  0.17952 & 99.990 & 99.990 & 19.831 & 99.990 & 99.990 \\
J033446.2+001710 &  53.71614 &   0.25407 &  0.32789 &  0.29340 & 99.990 & 20.378 & 18.895 & 18.209 & 18.020 \\
J035416.9-001003 &  58.52259 &  -0.15527 &  0.21438 &  0.20863 & 99.990 & 20.755 & 19.076 & 18.419 & 17.936 \\
J213340.8-003841 & 323.41968 &  -0.59219 & 99.99000 &  0.24568 & 99.990 & 22.762 & 21.722 & 21.430 & 99.990 \\
J221211.0-000833 & 333.02767 &  -0.10111 &  0.36528 &  0.36263 & 99.990 & 20.927 & 18.919 & 18.194 & 17.981 \\
J221422.1+004712 & 333.59307 &   0.78494 &  0.32024 &  0.34345 & 99.990 & 19.605 & 17.856 & 17.258 & 16.945 \\
J221449.2+004707 & 333.70879 &   0.74090 & 99.99000 &  0.35645 & 99.990 & 23.189 & 22.581 & 21.985 & 21.409 \\
J221722.9-001013 & 334.35941 &  -0.17238 &  0.33220 &  0.33786 & 99.990 & 20.406 & 18.675 & 18.013 & 17.525 \\
J222144.0-005306 & 335.42355 &  -0.88732 &  0.33658 &  0.39220 & 99.990 & 21.905 & 20.045 & 19.410 & 19.083 \\
J232540.3+001447 & 351.42001 &   0.24385 & 99.99000 &  0.74049 & 99.990 & 23.706 & 23.087 & 22.301 & 99.990 \\
J232613.8+000706 & 351.58764 &   0.11234 &  0.42613 &  0.42201 & 99.990 & 21.114 & 19.284 & 18.539 & 18.382 \\
J232742.1+001406 & 351.92704 &   0.22466 &  0.44513 &  0.44309 & 99.990 & 22.439 & 20.476 & 19.492 & 19.364 \\
J232809.0+001116 & 352.01612 &   0.15310 &  0.28008 &  0.29981 & 21.943 & 20.707 & 19.847 & 19.497 & 19.188 \\
J232925.6+000554 & 352.35567 &   0.09943 &  0.40193 &  0.38233 & 99.990 & 21.017 & 18.961 & 18.170 & 18.104 \\
J233138.1+000738 & 352.90866 &   0.12840 &  0.22382 &  0.22896 & 22.171 & 18.802 & 17.398 & 16.746 & 16.720 \\
J233328.1-000123 & 353.36624 &  -0.02276 &  0.51201 &  0.52293 & 99.990 & 21.227 & 19.387 & 18.469 & 18.099 \\
\hline
\end{tabular}
\label{tab:BCGs}
\end{table*}



\section{Schechter fit parameters for individual clusters}

The parameters of the Schechter fits up to the 80\% completeness level
are given in Table~\ref{tab:Sch_param_all} for all the analysed
clusters for which the GLFs converge. To make this table more
readable, we also choose to give the $M^*$ and $\alpha$ parameters, but
not the normalisation parameter $\Phi$, which is not very informative,
since it is usually not well constrained. All the clusters except one
(noted with an asterisk) are included in the GLF stacks in mass.  The
clusters indicated with a $z$ superscript are those included in the
GLF stacks in redshift.


\onecolumn
\voffset -2truecm
\begin{landscape} 
\begin{centering}
\begin{table*}
\scriptsize
\caption{Schechter parameters $M^*$ and $\alpha$ for the 36 individual clusters for which the fits have converged in at least one band.}
\begin{tabular}{|l|l|r|r|c|c|c|c|c|c|c|c|c|c|}
\hline
Cluster (3XMM)   & z     &R$_{200}$& M$_{500}$ & $M^*_u$ & $\alpha_u$ & $M^*_g$ & $\alpha_g$ & $M^*_r$ & $\alpha_r$ & $M^*_i$ & $\alpha_i$ & $M^*_z$ & $\alpha_z$ \\
                 &       & (kpc)  & ($10^{13}$~M$_\odot$) &&&&&&&&&& \\ 
\hline
\hline
J001115.5+005152 &0.3622 &729.8 &4.78& & & & & & &$-22.8\pm 2.0$ &$-1.26\pm 0.51$ &$-23.2\pm 2.6$ &$-0.97\pm 0.72$ \\
J001737.3-005240 &0.2141 &1120.1&14.67 &$-19.8\pm1.6$&$-1.14\pm 1.05$ & $-21.0\pm 0.04$&$-0.54\pm 0.28$ & $-21.7\pm 0.4$&$-0.67\pm 0.24$ &$-21.9\pm 0.4$ &$-0.65\pm 0.22$ &$-21.9\pm 0.4$ &$-0.28\pm 0.33$ \\
J002223.3+001201 &0.2789 &708.0 &3.98  & & & & & & &$-23.6\pm 7.6$ &$-1.09\pm 0.79$ & &  \\
J002314.4+001200 &0.2597 &779.1 &5.19  & & &$-21.4\pm 1.0$ &$-0.89\pm 0.37$ &$-22.2\pm 1.2$ &$-1.03\pm 0.30$ &$-22.2\pm 0.9$ &$-0.95\pm 0.29$ &$-23.4\pm 1.7$ &$-1.16\pm 0.35$  \\
J002928.6-001250 &0.06   &709.1 &3.18  &$-18.5\pm 0.6$ &$-1.10\pm 0.30$ &$-20.5\pm 0.5$ &$-1.05\pm 0.16$ &$-21.46\pm 0.4$ &$1.00\pm 0.15$ &$-21.0\pm 0.4$ &$-0.88\pm 0.19$ &$-21.3\pm 0.5$ &$-0.85\pm 0.26$  \\
J003838.0+004351 &0.6955 &962.8 &16.29 & & & & & & &$-23.7\pm 1.6$ &$-1.34\pm 0.37$ & &  \\
J003840.3+004747 &0.5553 &804.6 &8.04  & & & & & & &$-24.7\pm 7.8$ &$-1.50\pm 0.68$ &$-23.3\pm 1.5$ &$-0.94\pm 0.77$  \\
J003922.4+004809 &0.4145 &958.6 &11.52 & & & & &$-23.1\pm 3.2$ &$-1.45\pm 0.55$ &$-22.6\pm 1.5$ &$-0.72\pm 0.48$ & &  \\
J003942.2+004533 &0.4156 &851.0 &8.07  & & & & & & &$-23.9\pm 2.7$ &$-1.22\pm 0.48$ & &  \\
J004252.5+004300 &0.2697 &800.8 &8.36  &$-23.0\pm 12.1$ &$-1.59\pm 1.36$ &$-21.7\pm 1.9$ &$-0.76\pm 0.48$ &$-22.1\pm 1.14$ &$-0.76\pm 0.34$ &$-22.4\pm 1.2$ &$-0.66\pm 0.43$ &$-23.3\pm 2.1$ &$-0.81\pm 0.51$  \\
J004334.1+010107 &0.2    &910.0 &7.67  &$-26.0\pm 13.8$ &$-1.93\pm 0.33$ &$-21.58\pm 1.1$ &$-0.84\pm 0.45$ &$-21.85\pm 0.75$ &$-1.04\pm 0.28$ &$-22.0\pm 0.7$ &$-0.87\pm 0.37$ &$-23.0\pm 1.3$ &$-1.03\pm 0.47$  \\
J004401.4+000644 &0.2187 &951.9 &9.05  & & &$-21.8\pm 1.5$ &$-1.14\pm 0.26$ &$-23.8\pm 3.3$ &$-1.24\pm 0.18$ &$-22.8\pm 0.6$ &$-1.14\pm 0.13$ & &  \\
J005546.1+003839 &0.0665 &589.6 &1.84  & & & & & & & & &$-26.0\pm 9.9$ &$-1.10\pm 0.84$  \\
J010606.7+004925 &0.2564 &1026.9&11.84 &$-20.4\pm 1.9$ &$-1.84\pm 1.28$ &$-20.7\pm 0.5$ &$-0.82\pm 0.25$ &$-21.5\pm 0.6$ &$-0.89\pm 0.20$ &$-21.6\pm 0.6$ &$-0.72\pm 0.23$ &$-22.0\pm 0.8$ &$-0.75\pm 0.32$  \\
J012023.3-000444 &0.078  &903.6 &6.69  &$-20.3\pm 2.4$ &$-1.14\pm 0.60$ &$-21.4\pm 1.2$ &$-0.80\pm 0.46$ & & & & &$-23.2\pm 1.8$ & $-1.01\pm 0.40$ \\
J021012.6-001439 &0.2828 &791.3 &5.58  & & &$-24.0\pm 12.4$ &$-1.22\pm 0.79$ &$-26.00\pm 9.0$ &$-1.48\pm 0.25$ &$-26.0\pm 12.7$ &$-1.40\pm 0.26$ &$-25.1\pm 8.0$ &$-1.25\pm 0.85$  \\
J021045.8-002156 &0.31   &756.8 &5.03  & & &$-26.0\pm 9.9$ &$-1.12\pm 0.34$ &$-26.0\pm 13.5$ &$-1.40\pm 0.22$ &$-26.0\pm 12.1$ &$-1.12\pm 0.26$ &$-23.0\pm 1.7$ &$-0.85\pm 0.64$  \\
J022825.8+003203 &0.3952 &1035.8&14.21 & & &$-22.8\pm 4.6$ &$-1.69\pm 0.88$ &$-21.7\pm 0.9$ &$-0.80\pm 0.49$ &$-22.1\pm 0.9$ &$-1.16\pm 0.37$ &$-22.0\pm 0.6$ & $-0.56\pm 0.50$ \\
J025846.5+001219 &0.2589 &909.6 &8.25  & & & & & & &$-21.2\pm 1.0$ &$-0.63\pm 0.63$ &$-21.8\pm 1.2$ &$-0.85\pm 0.66$  \\
J025932.5+001353 &0.192  &1012.6&10.59 &$-20.1\pm 1.9$ &$-1.57\pm 0.95$ &$-21.5\pm 0.8$ &$-1.04\pm 0.25$ &$-21.9\pm 0.6$ &$-0.93\pm 0.25$ &$-23.4\pm 1.7$ &$-1.27\pm 0.20$ &$-22.8\pm 0.81$ &$-1.07\pm 0.32$  \\
J030205.6-000001 &0.65   &862.4 &11.09 & & & & &$-26.0\pm 8.0$ &$-1.21\pm 0.45$ & & & &  \\
J030212.1+001107 &0.6523 &658.1 &4.94  & & & & &$-26.0\pm 7.2$ &$-1.81\pm 0.53$ &$-26.0\pm 7.0$ &$-0.72\pm 0.37$ & &  \\
J030317.5+001245 &0.59   &801.3 &8.28  & & & & &$-26.0\pm 12.2$ &$-1.75\pm 0.34$ &$-23.6\pm 3.0$ &$-1.16\pm 0.78$ & &  \\
J030614.1-000540 &0.4249 &890.9 &9.36  & & & & &$-21.8\pm 1.9$ &$-1.33\pm 0.79$ & & & &  \\
J033446.2+001710 &0.3261 &943.6 &9.92  & & & & &$-23.4\pm 3.9$ &$-1.01\pm 0.63$ &$-23.5\pm 2.5$ &$-0.97\pm 0.48$ &$-23.2\pm 1.8$ &$-0.53\pm 0.78$  \\
J035416.9-001003 &0.21   &996.0 &10.27 & & &$-26.0\pm 11.1$ &$-1.62\pm 0.12$ &$-26.0\pm 13.7$ &$-1.60\pm 0.15$ &$-26.0\pm 12.8$ &$-1.43\pm 0.13$ &$-26.0\pm 14.0$ &$-1.07\pm 0.22$  \\
J221211.0-000833 &0.3643 &939.7 &10.24 & & &$-21.6\pm 1.4$ &$-1.70\pm 0.38$ &$-21.1\pm 1.2$ &$-1.12\pm 0.56$ &$-21.7\pm 1.0$ &$-1.14\pm 0.43$ &$-23.8\pm 2.2$ &$-1.62\pm 0.34$  \\
J221422.1+004712 &0.3202 &893.0 &8.36  & & &$-21.9\pm 1.1$ &$-1.06\pm 0.37$ &$-22.4\pm 1.0$ &$-1.14\pm 0.28$ &$-21.9\pm 0.8$ &$-0.73\pm 0.37$ &$-23.6\pm 1.7$ &$-1.25\pm 0.39$  \\
J221449.2+004707 &0.3171 &877.7 &7.91  & & &$-26.0\pm 14.0$ &$-1.84\pm 0.19$ &$-26.0\pm 13.9$ &$-1.61\pm 0.18$ &$-26.0\pm 8.8$ &$-1.49\pm 0.16$ & &  \\
J221722.9-001013 &0.3314 &896.4 &8.56  & & &$-25.8\pm 9.7$ &$-1.87\pm 0.40$ &$-22.2\pm 1.1$ &$-0.85\pm 0.43$ &$-21.9\pm 1.0$ &$-0.82\pm 0.52$ &$-23.7\pm 3.3$ &$-1.36\pm 0.67$  \\
J222144.0-005306 &0.3353 &896.9 &8.62  & & & & &$-22.6\pm 3.4$ &$-1.67\pm 0.50$ &$-25.8\pm 11.6$ &$-1.36\pm 0.52$ & &  \\
J232613.8+000706 &0.4261 &676.6 &3.91  & & & & &$-21.9\pm 1.0$ &$-0.54\pm 0.71$ & & &$-26.0\pm 13.9$ &$-1.22\pm 0.52$  \\
J232809.0+001116 &0.278  &831.0 &6.43  &$-25.8\pm 13.9$ &$-2.23\pm 0.71$ &$-26.0\pm 10.8$ &$-1.64\pm 0.30$ &$-26.0\pm 12.3$ &$-1.74\pm 0.15$ &$-26.0\pm 7.0$ &$-1.55\pm 0.17$ &$-26.0\pm 11.9$ &$-1.39\pm 0.25$  \\
J232925.6+000554 &0.4021 &709.0 &4.59  & & &$-25.8\pm 10.5$ &$-1.72\pm 0.39$ &$-23.1\pm 2.5$ &$-1.35\pm 0.55$ &$-22.1\pm 1.1$ &$-0.75\pm 0.61$ &$-22.47\pm 1.7$ &$-1.00\pm 0.97$  \\
J233138.1+000738 &0.2238 &777.6 &5.02  & & &$-25.8\pm 7.0$ &$-1.67\pm 0.20$ &$-26.0\pm 7.9$ &$-1.62\pm 0.24$ &$-26.0\pm 13.9$ &$-1.65\pm 0.26$ &$-26.0\pm 13.3$ &$-1.52\pm 0.32$  \\
J233328.1-000123 &0.512  &876.8 &9.32  & & & & & & & & &$-21.8\pm 9.0$ &$-0.58\pm 1.45$  \\
\hline
\label{tab:Sch_param_all}
\end{tabular}
\end{table*}
\end{centering}
\end{landscape}



\bsp	
\label{lastpage}
\end{document}